%% file: ms.tex
\documentclass[12pt,preprint]{aastex}
\begin{document}

\received{}
\revised{}
\accepted{}

\title{Mergers of close primordial binaries}
\author{N. Andronov, M. H. Pinsonneault, and D. M. Terndrup}

\affil{The Ohio State University, 
Department of Astronomy, Columbus, OH 43210}

\email{andronov, pinsono, terndrup@astronomy.ohio-state.edu}

\slugcomment{Draft in preparation for The Astrophysical Journal}
\shorttitle{Close binary mergers}
\shortauthors{Andronov et al.}

\keywords{stars: binaries, stars: blue stragglers, stars:evolution}

\begin{abstract}
We study the production of main sequence mergers of tidally-synchronized 
primordial short-period binaries. The principal ingredients of our calculation are the 
angular momentum loss rates inferred from the spindown of open cluster stars 
and the distribution of binary properties in young open clusters. 
We compare our results with the expected number of systems that 
experience mass transfer in post-main sequence phases of evolution 
and compute the uncertainties in the theoretical predictions. 
We estimate that main-sequence mergers can account for 
the observed number of single blue stragglers in M67. Applied to the 
blue straggler population, this implies that such mergers are 
responsible for about one quarter of the population of halo blue metal poor 
stars, and at least one third of the 
blue stragglers in open clusters for systems older than 1 Gyr.
The observed trends as a function of age are consistent with a saturated 
angular momentum loss rate for rapidly rotating tidally synchronized systems.
The predicted number of blue stragglers from main sequence mergers alone is 
comparable to the number observed in globular clusters, indicating that the 
net effect of dynamical interactions in dense stellar environments is to 
reduce rather than increase the blue straggler population.
A population of subturnoff mergers of order 3-4\% of the 
upper main sequence population is also predicted for stars 
older than 4 Gyr, which is roughly comparable to the small population of 
highly Li-depleted halo dwarfs.  Other observational tests are 
discussed. 
\end{abstract}

\clearpage
%\tableofcontents
%\clearpage

\section{Introduction}
About half of all stars are found in binaries, and about 30\% of binary 
stars will interact at some point in their lifetime. Interacting binaries 
can therefore represent a significant fraction of all stars, and the 
consequences of interactions can be important for understanding stellar 
populations. In fact, studies of star clusters often reveal 
examples of ``stellar anomalies'' not explainable by single-star evolution,
but which can be attributed to interacting binaries \citep{bailyn95}.

Nature provides a variety of ways that binary stars can interact, with many 
possibilities depending upon the relative masses and evolutionary states 
of the components \citep[see][]{paczynski71,il93}. In particular, 
binary interaction or merger may be responsible for the 
existence of blue stragglers (BSs). They are stars which appear 
on the color-magnitude diagram of a cluster on a natural extension of 
the main sequence above the turn-off point. 
The existence of BSs was discovered by \citet{sandage53} in the globular 
cluster M3. Over the years blue stragglers have been identified in many globular and 
open clusters, as reviewed by \citet{stryker93}, and they are numerous in some systems; 
the number of BSs found in individual globular clusters ranges from 
10 to 400 \citep[]{bailyn95,dpd04}.

Binary mergers are by no means the only way to produce blue stragglers, and not 
all mergers would manifest themselves as blue stragglers. Many mechanisms for 
the origin of BSs have been proposed; a summary can be found in \citet{abt85}. 
There are two general categories: those in which
BSs were formed later than the general stellar 
population of a cluster, or those in which the some mechanism prolongs the 
main sequence lifetime of a minority of stars. In the former category are 
proposals that blue stragglers may arise from stellar mergers or collisions 
(see \citet{dpd04} for a discussion). In sufficiently dense stellar environments, 
collisions between stars can be an important production mechanism.  

Alternatively or concurrently, binaries with a wide range of orbital periods
can either merge or experience mass transfer. Binary systems with 
orbital periods above 5 days but with initial separations below 3 AU may overfill their 
Roche lobes after one of the members leaves the main sequence. The ensuing 
mass transfer would produce a variety of unusual stellar objects such as 
common envelope systems, X-ray binaries, binary pulsars, and Blue Stragglers 
\citep{bailyn95}. Shorter period systems, however, can even interact on the main 
sequence, and because of angular momentum loss complete mergers are more likely
(see section 3.3). 

There are two main mechanisms that can cause angular momentum loss in close binaries. 
Binary orbits decay from gravitational radiation. In addition, angular momentum 
loss from the magnetized winds of late-type stars \citep[e.g.,][]{wd67}
results in a slowing of stellar rotation rates with time \citep{skumanich72}. 
In tidally synchronized systems of short initial period, such angular momentum 
loss would lead to a reduction of the size of the orbit and produce 
contact systems or eventual mergers. Although short-period binaries are 
rare in the field, they are substantially more common in young open 
clusters, which supports the idea that such systems eventually merge. 
For example, 11\% of Hyades binaries have orbital periods between 0.5 and 5 days \citep{dm91}.  
The ``collision'' and ``merger'' categories are not completely distinct; close 
stellar encounters can form tight binaries which subsequently merge.
Following the conventional naming \citep[e.g.,][]{dpd04}, 
we would call the first type dynamical BSs, and the second type primordial BSs.

Although most work on the origin of BSs has focused on mergers,
recent theoretical work on rotational mixing has revived the possibility
that some BSs are ``normal'' stars that have had their lifetime increased relative
to other stars of the same initial mass. The most promising candidate 
for main sequence lifetime extension is mixing induced by rapid rotation 
in some massive stars \citep[see][for a review]{mm00}.
This is probably not a major channel for producing lower mass blue stragglers 
(below about 1.5 solar masses) because extensive angular momentum loss 
in lower main sequence stars makes rotational mixing relatively inefficient \citep{pinsonneault95}. 
However, it may be important in explaining the presence of blue stragglers 
in open clusters too young for the effective production of dynamical 
or primordial BSs \citep[]{mermilliod82}.

Determining the relative importance of these
different channels can provide important insights into a number of related 
fields in stellar astrophysics. Are star clusters valid templates for stellar 
population studies, or do dynamical effects significantly modify their global 
properties relative to field stars?  How important are blue stragglers for 
the integrated light of stellar populations?  In addition, the production 
rate for stellar mergers can potentially be used as a test of the angular 
momentum loss rate for tidally synchronized binaries. 

There may also be interesting consequences of stellar mergers for our understanding of surface 
abundance anomalies. There is no {\it a priori} reason why all stellar mergers or mass transfer 
events must manifest themselves as blue stragglers.  Some merger products 
will appear below the main sequence turnoff.
Such objects would be slightly less evolved than their single counterparts, 
but might have unusual rotation rates and surface abundances.  There is an 
interesting subpopulation of highly over-depleted stars found in lithium 
studies in open clusters \citep[e.g.,][]{thorburn93} and halo field samples 
\citep{thorburn94}. Because lithium is destroyed at low temperatures in 
stellar interiors, either internal mixing or mergers could produce low surface lithium abundances. 
\citet{ryan01} have argued that the highly lithium depleted halo stars may have experienced mass 
transfer from a companion, and found that some such stars close to the 
turnoff rotate significantly more rapidly than their lithium-normal counterparts. 
The frequency of mergers, needs to be quantified, however, to distinguish between 
the merger and mixing hypotheses.

In this paper we use recent prescriptions of angular momentum loss
to compute the rates of mergers of main-sequence tidally synchronized binaries 
with short initial periods ($\lesssim 5$ days). Our paper 
is organized as follows.  We describe recent observational advances in 
the study of BSs and potential subturnoff merger products in section 2.1. 
In section 2.2 we outline our motivation for studying BSs formed from primordial 
binaries in open clusters.  Section 3 gives an overview of our model, then 
we proceed with discussion of observational data and selection in section 4. 
We present our results in section 5, and section 6 contains our summary and conclusions.

\section{Background on the population of blue stragglers}
\subsection{Stragglers in different environments}
A considerable amount of ground-based data on blue stragglers in globular clusters 
has been collected \citep[e.g.,][]{stryker93}. With the use of HST 
the amount and quality of the data has increased, which has provided 
additional clues about the origins of BSs.

Most globular clusters have a centrally concentrated radial distribution of Blue 
Stragglers \citep[e.g.,][]{stryker93}. There are two known exceptions to date. 
The first is M3; it has been studied both from the ground \citep[]{fer93} 
and subsequently from space \citet[]{guh94}, and it was found that it has a peculiar
bimodal radial distribution of BSs. \citet[]{fer04} also found that 47 Tuc   
has a bimodal radial distribution of BSs similar to M3. The existence of two radial
peaks is suggestive that environmental effects are important. It is plausible 
that dynamical effects could be important for BSs produced in
globular clusters \citep[e.g.,][]{bailyn92}. In high density 
stellar environments (i.e., globular cluster cores) collisional mergers 
maybe created. The other production mechanism (MS mergers and post-MS mass transfer
in primordial binaries) would dominate when the stellar density is low 
\citep[e.g.,][]{mccrea64}. This would be the case in the outer parts 
of globular clusters, open clusters and 
in the Galactic halo. This idea is somewhat supported by 
differences in the luminosity functions of BSs in the inner 
and outer regions of M3 \citep[]{bp95}.

The similarities in the M3 and 47 Tuc BS populations 
are, however, problematic for models that rely on collisions alone, because the 
two clusters are very different dynamically \citep[]{fer04}. 
The central density of 47 Tuc is about 40 times larger 
than that of M3 \citep[see][]{pm93}; also, other tracers of 
dynamical interactions, such as X-ray binaries are relatively 
abundant in 47 Tuc but not in M3 \citep[]{fer01}. \citet{fer03} argued that the BSs 
frequency cannot be parameterized uniquely by the collision rate. This 
result has been strengthened to some extent recently by 
\citet{piotto04}, who summarized data on about 3000 BSs 
in 56 globular clusters and found that there is no statistically 
significant correlation between BSs frequency and the collision 
rate of stars in the cluster cores. 

Blue Stragglers have also been found in open clusters;  for a summary,
see \citet{al95}. Open clusters have much lower central stellar densities 
than globular clusters do, and therefore dynamical effects are less important
in open clusters than in globular clusters. Field populations provide another
potential test of environmental effects.
Blue stragglers are more difficult 
to discover in the field, which contains stars with a wide range of ages. 
It is possible, however, to look in the Galactic halo for field 
analogs of the blue stragglers seen in clusters. High velocity, blue metal 
poor (BMP) stars in the Galactic halo have been studied by \citet{ps00} 
and by \citet{carn01}. Both groups found that a large fraction of 
BMP stars (50 - 60\%) are members of binaries with orbital 
periods from half year up to a few years. The unseen companions are 
consistent with being white dwarfs of mass 
$\approx 0.55 M_\odot$. \citet{ps00} found that such systems are about four 
times more abundant than HB stars in the same field (in globulars, 
by comparison, the number of blue stragglers and HB stars are approximately 
equal). They have associated BMP stars with Blue Stragglers, and argued that 
mass transfer from an expanding giant on to a main sequence star would 
be the dominant mechanism of formation of BSs in low density stellar environments.

Most of the arguments above are indirect. However, there is recent 
observational support for the idea that some close MS binaries become 
contact systems and therefore might be associated with BSs (by their position on a 
color-magnitude diagram) or produce single-stars above turn-off, if coalescence 
is sufficiently rapid. Studying the BSs population of NGC 5466, \citet{mateo90}
have discovered that three of the BSs are close binaries, two of them being 
full contact systems. Eclipsing and/or contact binaries have later been found in NGC4372 
\citep{kk93}, M71 \citep{ym94}, NGC6791 \citep{rkh96}, NGC2354 \citep{la96},
NGC 6752 \citep{thompson99}, and NGC 6362 \citep{mkk99}.
Finally, \citet{vdb01} studied a blue straggler in the old open cluster 
M67 and found that it is a triple system with inner 
binary having a mass more than twice of the cluster's turn-off.
This implies that at least one of the stars has a mass greater than the 
turn-off mass and is therefore a BS. In addition, the third star is also a BS. 
BSs production can therefore be linked firmly to at least some close binary systems.

\subsection{Prior theory}
The numerous theoretical possibilities  for BSs production have 
naturally stimulated a number of theoretical papers. Some have 
included elaborate dynamical and evolutionary calculations 
for specific interesting systems that we do not include.
A good example is the work of \citet{hurley01}. 
They combined dynamical N-body simulations with simplified models 
of stellar and binary evolution \citep{hurley00}.
Dynamical effects can increase the production of BSs, explain the existence 
of those in binaries, and quantify the number 
of other stellar exotica observed (such as RS CVns and post-mass transfer BSs). 
The authors claimed that binary evolution alone can not account
for the number of observed BSs in open clusters, or the binary properties of these BSs.
These models are potentially much more powerful than the relatively simple 
treatment of binary population used in this work. However, they suffer 
from difficult-to-quantify effects related to the choice of initial 
conditions (including the spatial distribution
of stars) and their impact on the subsequent dynamical phenomena.
There are also numerical challenges such integrating binary 
trajectories during encounters with sufficient precision. 

A simple treatment of binary population seems to be much
more transparent as a test for the initial properties of the stellar population
and magnetic braking for a larger sample of systems. 
It is also straightforward to estimate the theoretical
uncertainties involved. Even if dynamical phenomena are required in M67, such
massive open clusters are rather unusual
in the galactic open cluster population. Numerous low mass clusters, where 
dynamical effects can be neglected, provide a wealth of data on BSs whose 
formation should not be influenced by dynamical effects. A comparison of the 
number of open cluster BSs with the predictions of our model can in 
principle provide stronger constraints than the study of one cluster 
on the initial period distribution, mass function, and angular momentum 
loss rate of close binaries.

In addition, there have been fewer papers devoted to computing the MS binary merger
rates, which is our primary concern. \citet{vilhu82} and 
\citet{stepien95} concluded that contact systems can form
from a population of close binaries (with period below about 4-6 days), and that
orbital decay of primordial binaries is sufficient to explain the existence of 
contact systems. Their models were restricted to the study of small set 
of initial binary parameters. 
The theoretical uncertainties were also only roughly estimated. 
We take the next logical step, assuming that the 
formation of a rapidly rotating single star will result from the formation of 
such contact binaries. Then, given the lifetime of a merged product, we 
can compute the abundance of such mergers and compare them to much more 
numerous population of BSs in open clusters.

Our work therefore extends prior studies and in a few aspects differs from previous models.
We use a much more realistic model of angular momentum loss, motivated by 
recent studies on the spindown of low mass MS stars. 
We model merger production from a distribution of masses, rather than
calculating a few representative mass ratios. We also investigate 
the effects of choosing different stellar mass functions and
initial period distribution of binaries, supported by the latest 
available empirical data. We calculate the total number of mergers, 
taking into account the lifetime of a 
merged product for different assumptions about its physics. A constant 
lifetime for the contact binary phase was used in \citet{vilhu82}. We also 
compute the expected number of subturnoff mergers. 
We compare predictions of our model with observed BSs in open clusters with a variety
of ages, where the statistics are better than those for the sparse population 
of contact MS binaries. We investigate the impact of observational selection criteria. 

\section{The model}
Observational studies of BSs usually concentrate on 
inferring the relative number of blue stragglers relative 
with respect to the number on the upper main sequence or 
the horizontal branch population.  The theoretical models
presented here, therefore, reduce to the problem of 
determining the rate of merger production relative to
the lifetimes in the other evolutionary states.  We begin with an 
overview of the ingredients of the theoretical
models, which can be divided into three broad categories. We then follow with 
detailed discussions of the individual components of the models.

The initial conditions (distribution of primary masses, relative primary
and secondary masses, binary fraction, and the starting binary period 
distribution) are considered
in section 3.1, while section 3.2 discusses the rate mergers take
place;  this depends primarily on the rate of angular momentum
loss and the time scale for tidal synchronization. Finally, the physics 
of the merger process itself will have a significant impact on the lifetimes 
and predicted positions of merger products in the HR diagram, as
discussed in section 3.3. In section 3.4 
we compute the relative number of BS stars and of sub-turnoff
mergers.

\subsection{Properties of stellar population}
We first assume that all stars in an open cluster have 
the same age and metallicity. Then we assume that the mass 
function of stars, binary fraction, and binary period distribution are independent of 
the age, mass, metallicity of a cluster. Evaporation of
stars via dynamical effects \citep[e.g.,][]{fuente95,kim99} is not included 
in the model, since evaporation will mainly remove low-mass stars. 
Such objects do not have the combined mass to appear as BSs after merging, and because BSs
counts are expressed relative to the upper MS population, the loss of low mass stars does 
not impact the observatinal interpretation. Finaly, we assume 
that collisions do not alter the binary period distribution.  

Each model cluster initially consists of $N$ stars.  Let $N_1$ be 
the number of single stars and $N_2$ the number of binary sytems; then $N = N_1 + 2N_2$.  

The binary fraction may be characterized by 
\[
\epsilon \equiv {
        N_2
   \over 
        { N_1 + N_2 }
                }.
\] 
Typically, $\epsilon$ has the value $0.2 - 0.5$ 
(e.g., Duquennoy \& Mayor, 1991).   
In terms of the binary fraction,
the number of binaries is then given by
\[
N_2 = N\left(\frac{\epsilon}{1+\epsilon}\right),
\]
and the number of single stars is
\[
N_1 = N\left(\frac{1-\epsilon}{1+\epsilon}\right).
\]

We model the binary population by a distribution 
function over primary mass $m_1$, secondary mass
$m_2$ and orbital period $P$;  here $m_1$ and $m_2$ are in solar
units, and $m_1 \geq m_2$. The number of binaries in the 
interval between $(m_1, m_2, P)$
and $(m_1 + dm_1, m_2 + dm_2, P+dP)$ is given by
\[
dN = F(m_1,m_2,P) dm_1 dm_2 dP.
\]
We assume that the distribution function 
$F(m_1,m_2,P)$ can be decomposed as a product 
of distribution functions over $m_1$, $m_2$ and $P$, namely
\[
F(m_1,m_2,P) = \xi_1(m_1) \xi_2(m_2)f(P),
\]
where $f(P)$ is an initial period distribution, and $\xi_1(m_1)$
and $\xi_2(m_2)$ are mass functions for the primary 
and the secondary stars, respectively, which are not necessarily identical. 

The mass distribution of single stars is generated from
a variety of mass functions (MFs).  Our base case is
the \citet{ms79} function, which can be expressed
as
\[
\xi(m) \propto m^{-2-0.5\log(m)}.
\]
In addition, we also explored the effect of including
an IMF that is much steeper at lower masses, namely
the \cite{salpeter55} function
\[
\xi(m) \propto m^{-2.35},
\]
and a shallower IMF which characterizes the distribution
of stars in the open cluster M35 as obtained by
\citet{barrado01}. Recent observations of the Pleiades down to the hydrogen burning boundary 
showed that the cluster also has a mass function similar to that of M35
\citep{moraux03}. Our examination of their data
yields approximately
\[
\xi(m) \propto m^{-1.5-\log(m)}.
\]
These three IMFs are shown in Figure 1, all normalized 
at the solar mass. We treat the Salpeter and  M35 mass functions 
as limiting cases. We will demonstrate that model is mostly sensitive to 
the slope of the IMF in the interval $0.4 - 1.0M_\odot$, so our results
are relatively insensitive to the primary star IMF. 

We assume that the masses of the primary stars are distributed according
to the same IMF as the single stars. For the secondaries,
we test two possibilities: that they have the same IMF as the primaries, or that
they have a flat mass function $\xi_2(m_2) = {\rm const}$. The second choice
is dictated by accumulating observational evidence that 
the relative masses of close binary stars are more equal than would be expected
if they were drawn from the same IMF as the primary masses \citep[]{abt04}. 

For the binary period distribution we use an empirical fit to the data summarized by \citet{dm91} 
for G-dwarfs in the solar neighborhood, which is characterized by a Gaussian in $\log P$. 
The period distribution of dwarf stars in the Hyades \citep{griffin85} is consistent 
with this form for periods over 10 days, but has more binaries with shorter periods.
\citet{dm91} speculate that since the Hyades field is significantly  younger than the field 
population (the age of Hyades is about 600 Myrs), the excess at short periods is a 
characteristic feature of young populations, which evolves away with time
due to the coalescence of short period binaries. We will demonstrate below 
that short-period systems must be frequent relative to the field distribution in
order produce a significant number of BSs from mergers. 

We therefore assume that period distribution of binaries is
given by
\[
\frac{dN}{d(log P)} \propto
\exp{\left[  -\frac{ (logP-\alpha)^2}{2 \sigma_1^2}\right]}
+\frac{2}{3}\exp{\left[  -\frac{ (logP)^2}{2 \sigma_2^2}\right],}
\]
where $P$ is orbital period in days.  The first term 
describes the field distribution from Duquennoy \& Mayor (1991),
who derived $\alpha = 4.8$ and $\sigma_1 = 2.3$;  we 
adopted these values in our model. 
The second term describes short period systems which we
assume exist in young stellar systems.  An inspection 
of the Hyades data from \citet{griffin85} yielded
$\sigma_2 \approx 0.6$ and gave us the $2/3$ term characterizing
the relative normalization of the distributions of short 
and long periods. We truncate this distribution at
short periods, seting the boundary at either 0.5 days 
or the initial period the binary would immediately be 
a contact system, whichever is longer.

\subsection{Orbital decay of close binaries}
\subsubsection{Tidal synchronization}
Loss of orbital angular momentum will result in a reduction
of the orbital size in binaries, leading eventually to
contact and merger. There are two mechanisms for angular 
momentum loss.  The first is gravitational radiation
\citep[e.g.,][]{ll62}, which is inefficient for periods 
above about 0.5 days.  The second is loss by magnetized winds 
from main sequence stars, discussed below.  

Each binary is assumed to begin already in tidal
synchronization. This assumption is valid if the time
scale for synchronization is shorter than any other
time scale involved in the problem. According to
\citet{zahn77}, the time to synchronization at
any given orbital period $P$ may be approximated by 
\[
\tau[yr]\approx 10^{4} \left[
{ {1 + q} \over {2q} }
\right]^2 P^4,
\] 
where $P$ is in days and $q = m_2 / m_1$. 

As will be discussed below, our prescription for angular momentum loss means that 
the longest initial orbital period which would produce a contact binary in 10 Gyr is about 5-6 days;
this is valid mainly for cases where $m_2 \approx m_1 > M_{\odot}$. In these circumstances, 
$\tau \approx 10^7$ yr. If the secondary is considerably less massive than the primary, 
the synchronization time for the primary increases significantly, proportional to the square of 
the mass ratio $q$. For $m_1 = 1.0$ and $m_2 = 0.1$, $\tau$ is about 1 Gyr for
an initial period of 6 days. In this case, the angular momentum loss from 
the primary would not be removed from the orbital angular momentum and instead the primary would
simply spin down as it loses angular momentum. To take
this effect into account, we introduce a synchronization parameter, as discussed below.

In the case where both stars are in synchrony with the orbit, the total angular 
momentum in the binary system is given by
\begin{equation}
J = M_{\odot}^{5/3} G^{2/3} m_1 m_2 m^{-1/3}\omega^{-1/3}+
[I(m_1)+I(m_2)]\omega, \label{simplej}
\end{equation}
Here $M_\odot$ is the solar mass, $G$ the gravitational
constant, $m = (m_1 + m_2)$, $\omega$ is the angular rate
of the orbit, and $I_1$ and $I_2$ are the stellar moments of inertia
for the primary and secondary, respectively. The first term
represents the orbital angular momentum, while the term in brackets 
is the rotational angular momentum of two stars.

To account for the longer time scale for synchronization
when $m_1 \gg m_2$, we modify equation \ref{simplej} by
assuming that the secondary is is always tidally locked, 
but the primary is locked only when the secondary is
sufficiently massive. We define a parameter $q_{\rm lock}
= m_2 / m_1$ to represent the binary mass ratio above
which the primary will be locked. Then the angular
momentum of the binary system, is
\begin{equation}
J = M_{\odot}^{5/3} G^{2/3} m_1 m_2 m^{-1/3}\omega^{-1/3}+
[\chi I(m_1)+I(m_2)]\omega, \label{jlaw}
\end{equation}
where $\chi = 1$ for $q > q_{lock}$ and $\chi = 0$
otherwise. The value of $q_{lock}$ should, formally, be some
function of age of the cluster but for simplicity we
set it to be a constant in the range
between 0.1 and 0.5. As shown below, the rate of
mergers is only slightly sensitive to 
the adopted $\chi$, mainly because binaries with $q \approx 1$
contribute most to the production of BSs.

\subsubsection {Prescription for angular momentum loss}
For the magnetic braking, we use an empirical prescription 
derived from studies of the rotation rates of stars in
young open clusters \citep{kmc95,kpbs97,terndrup00}. 
The total angular momentum extracted from the orbit for a 
fully synchronized system per unit time can be written as
\begin{equation}
\dot{J}_{\rm tot}(m_1,m_2,\omega)=
\dot{J}_{\rm wind}(m_2,\omega)+
\dot{J}_{\rm wind}(m_1,\omega). \label{jdot}
\end{equation}
The empirical data requires a mass dependent
saturation of the loss rate compared to the
\citet{skumanich72} law which simply scales as  $\omega^3$. This
is usually written as \footnote{See \citet{ap04} for an 
application of this braking law to the evolution 
of cataclysmic variables.}
\begin{equation}
\dot{J}_{wind} = -K_w \left(\frac{r}{m}\right)^{1/2} \times
\left\{
\begin{array}{ll}
\omega^3                & \omega \leq \omega_{crit}\\
\omega \omega^2_{crit}  & \omega > \omega_{crit} \label{losslaw}
\end{array}
\right.
\end{equation}
The saturation threshold $\omega_{crit}$ is the critical 
angular speed at which the magnetic field saturates,
and is a function of mass. We chose the same mass dependence 
as \citet{ap04}. $K_w$ is a constant calibrated 
to reproduce the solar rotation at the age of the Sun,
and has the value $2.59 \times 10^{47}$ g cm s$^2$ 
\citep[see][]{kpbs97}. In addition, we also investigated 
the case where dynamo saturation was neglected, i.e.,
$\dot{J} \propto \omega^3$. 

A main sequence star can generate a magnetic field only 
if it has a sufficiently thick convective envelope
\citep{dl78}. Hotter stars do not experience 
magnetic braking, and for spectral types type earlier 
than F8 the average rotational velocity 
is independent of age \citep{wh87}. We model this
by setting a parameter $M_{c}$, above which we
take $\dot{J}_{wind} = 0$. The value of $M_c$ was 
taken in the range between 1.2 and $1.4 M_{\odot}$.
The lower bound corresponds to the mass below which rotation 
decreases with time on the MS. The upper bound allows for the 
possibility that intermediate mass stars may be losing internal 
angular momentum while their surface rates are constant.
We differ in this aspect from earlier studies \citep[][for example]{stepien95}
that chose a much higher mass threshold (up to 3$M_\odot$) in order to produce 
BSs in young systems. In our view there is no necessity for such a high
mass threshold. Massive BSs are more likely to be produced by other mechanisms,
such as rotational mixing or accretion from post-MS companion.

Now we can calculate the initial period from which a binary with 
masses $m_1$ and $m_2$ and an angular momentum loss rate given 
by equation \ref{jdot} will spiral in to the point when one of the stars overfills its 
Roche lobe. Equations \ref{jlaw} and \ref{jdot} together can be written as
\begin{equation}
\left\{
\begin{array}{l}
J = A \omega^{-1/3}+B \omega,  \\
\dot{J} = \dot{J}_{\rm wind}(m_2,\omega) 
        + \chi \dot{J}_{\rm wind}(m_1,\omega) \label{jterms}\\
\end{array}
\right.
\end{equation}
where $A$ and $B$ are constants determined by the 
masses of the stars.  These are given by
\[
\begin{array}{l}
A= { {m_1 m_2} \over {m^{1/3}} }
   M_{\odot}^{5/3} G^{2/3}, \\
B = \chi I(m_1) + I(m_2). \\
\end{array}
\]
We take the time derivative of the first equation of system \ref{jterms} and
set it equal to the second equation, which yelds
\begin{equation}
\dot{\omega}
\left(B - \frac{1}{3} A \omega^{-4/3}\right) =  \dot{J}_{\rm tot}(m_1,m_2,\omega).
\end{equation}
Then the time required to form a contact binary
$\tau_x$ can be found by integrating this equation backwards 
in time, starting from the contact period at which one of the 
stars overfills its Roche lobe.  If $\omega_x$ is the spin rate
initially, and $\omega_c$ is the rate at contact, then
\begin{equation}
\tau_{x}(m_1,m_2) = - \int^{\omega_{x}}_{\omega_{c}}
\frac{B-\frac{1}{3}A\omega^{-4/3}}{\dot{J}_{\rm tot}(m_1,m_2,\omega)}d\omega .
\end{equation}

The evolution of orbital period as a function of 
time for representative cases is shown 
in Figure 2 for a variety of binaries; it is analogous to 
Figure 2 from \citet{stepien95}. 
We examined three mass combinations: $(1+1)M_\odot$, 
$(1+0.6)M_\odot$, $(0.6+0.6)M_\odot$. 
The initial periods were 2, 4, and 6 days. Given either the 
\citet{stepien95} magnetic braking prescription (see his 
formula 14 and 15) or our method described above, none of the systems
would reach contact in 10 Gyr, with the initial 
period of 6 days. \citet{stepien95} tested magnetic braking rates with different 
efficiencies, some of which produced considerable orbital decay rates even for longer periods.
As a result, for example, in one of his models, a binary system $(0.6+0.6)M_\odot$ 
reached the contact in about 8 Gyr. We believe that recent observational 
data makes these choices unlikely.

Our results for shorter period systems differ with \citet{stepien95} because our 
braking law is calibrated on the observed spindown of open clusters stars
and includes mass-dependent saturation in the loss rate.  
As a consequence, angular momentum loss is less efficient than the ones 
considered by \citet{stepien95}. The mass dependence of our results is therefore considerably 
different for systems with an initial period of 2 days. Given the angular momentum 
loss model that we are using, a $(0.6+0.6)M_\odot$ binary system does not become a contact system
in the age of Galaxy. This is in marked contrast to the \citet{stepien95} result, for which 
such a binary reaches contact before the higher mass systems do.
In our model, the angular momentum loss rate decreases rapidly for decreasing mass at short periods.
At long periods, all binaries would be in the unsaturated regime and would have
angular momentum loss rates comparable to the earlier work represented by \citet{stepien95} 
because the mass dependence in this regime is weak.
 
\subsection{Mergers of contact systems}
In our model we assume that once a binary forms a contact
system, the two stars merge on a time scale that is short
compared to other scales of interest.  We designate a BS 
to be a single star formed in such merger, provided that
the combined mass is larger than the mass 
of the turn off of a cluster at the time the merger
takes place. 

The timescale for such coalescence is uncertain. Estimates based on 
observed frequencies of contact system in various environments range from $10^7 - 10^8$ yr 
\citep[e.g.,][]{vv79,ei89}, while some theoretical models of the merger process can yield 
times on the order of 10 Gyr \citep{mochnacki81}. Here, we adopt the 
empirically estimated values. 

Clearly, if the duration of the contact phase is long, it will result 
in fewer BSs in young open clusters. We can obtain an order of magnitude estimate of 
the contact phase lifetime from recent observations of contact binaries in the relatvely well
studied cluster NGC 188. \citet{kafka03} found 13 variables
with orbital periods below 2 days in the upper two V magnitudes of the
main sequence. Two out of 13 stars appear to be binaries that reached contact
during the post-MS expansion of the primary; both of them are on the turnoff
of the cluster, and their orbital periods are above 1.7 days.
Two more systems have long periods of about 0.9 days, so they should
be pre-contact MS binaries. We, therefore, are left with 9 binaries
that probably set an upper limit on the number of contact systems.
There are about 1450 stars in the 2 brightest magnitudes of the MS \citep[]{platais03}.
The fraction of contact binaries, therefore, relative to the number of stars
in this range is about 0.6\%. For old clusters the ratio of 
merged binaries to contact systems will be about 
the ratio of lifetimes in the BS phase to that in the contact phase. 
As can be seen from Figure 13, we predict
the fraction of mergers somewhere between 4\% and 5\%. This will set an upper
limit on the timescale of the contact phase to be somewhere between 12\% and 15\%
of the lifetime of the merged product. 

Once the stars merge, the product can be chemically homogenized or else 
the gradient in abundance caused by previous nuclear reactions can be preserved (i.e.,
the product has a helium-enriched core while the merger added essentially 
unprocessed material to the envelope). See \citet[]{bp95,lombardi95,lombardi96} 
for detailed discussion of the internal structure of mergers. 

The assumed nature of the merged product will affect the main sequence lifetime,
with unmixed products having lower MS lifetime than mixed
ones.  As a result, the number of BS stars would be reduced
if the mergers were fully mixed. 
For the case when the merged product is completely mixed, the subsequent 
lifetime in the BS phase $\tau_{\rm BS}$ is approximately given by the MS 
lifetime of a star with mass $(m_1 + m_2)$, and a starting hydrogen 
abundance $X$ that is determined by the prior nuclear evolution.
For the unmixed case, the lifetime of the merger product is approximately
$$
\tau_{\rm BS} \approx \tau_{\rm MS}(m_1+m_2)
\left(
1-\frac{\tau_{\rm MS}(m_1)}{\tau_{\rm MS}(m_2)}
\right)
$$
The factor in the brackets describes the maximum 
fraction of unprocessed hydrogen
in the center of the secondary star.

\subsection {Number of mergers}
The properties of primordial binaries is specified by a 
distribution in $(m_1,m_2,P)$ space.  In order to calculate 
the number of mergers which would form in the age of a cluster 
by coalescence, we need to integrate this distribution over a 
region of parameter space specified by several inequalities:

1) The initial orbital period of a binary should not 
be higher than the largest period which would lead to
the formation of a contact binary in the age of a cluster. 
It should also be longer than the contact period for a 
binary of given masses.  This condition is given by
$$
P_{\rm contact}(m_1,m_2)\leq P_{\rm init} \leq P_{\rm max}(\tau_{\rm cluster},m_1,m_2)
$$
where $P_{\rm max}$ is calculated using 
equation 7.

2) For MS mergers, the primary must be
on the main sequence when system reaches contact and merges.
Therefore $m_1 \leq m_{\rm to}(\tau_{\rm cluster})$.

3) The finite age of the merged products should be taken into account. 
At the time of an observation $\tau_{\rm cluster}$ the formed merger should still be 
on the main sequence. If the orbital decay time is too 
short, merger will go through all its main sequence evolution. In this case 
we will not see it as a BS. This is a lower limit on the 
initial period of binary star. It can be written as:
$$
\tau_{\rm orbital}(P_{\rm init} \rightarrow P_{\rm contact})+\tau_{\rm product}
\geq\tau_{\rm cluster}  
$$
from here: 
$$
\tau_{\rm orbital}(P_{\rm init} \rightarrow P_{\rm contact})\geq\tau_{\rm cluster}
-\tau_{\rm product}
$$
where $P_{\rm init}$ is initial period of binary.

4) For comparison of the predicted number of mergers with number of Blue Stragglers,
a proper selection criteria should be invoked. We count only those mergers that closely suit 
the selection criteria used in particular sample. Different observers
use different selection criteria and the selection criteria in our model must be 
modified accordingly. In particular we considered two possible 
selections. 

One is a simple luminosity cut; a merger product should have a luminosity at 
least 0.5 magnitudes greater than the turn-off luminosity of a cluster. This criteria 
closely resembles the one used in studies of globular clusters (Piotto et 
al 2004 for example). Another selection is based on a color cut; the 
effective temperature of a product should be higher than that of the 
turn off, but a product is allowed to be subluminous than the turn-off 
by no more than 1 magnitude. Such a theoretical selection resembles the one used in the
compilation of open cluster data by \citet[]{al95}. The results of 
adopting these different selection criteria are discussed in section 5.1.1.

As an example, we consider the selection criteria: $(m_1 + m_2 \ge m_{\rm to})$,
where $m_{\rm to}$ is the turnoff mass. This simplified form of 
selection is analogous to the luminosity cut in 
the unevolved population. Then, a region in the parameter space $(P,m_1,m_2)$ 
described above can be represented by a set of inequalities;
\begin{equation}
\left\{
\begin{array}{l}
m_2 \leq m_1 \leq m_{\rm to}(\tau_{\rm cluster})\leq m_1 + m_2,\\
P_{\rm contact}(m_1,m_2) \leq P_{\rm init},\\
\tau_{\rm cluster}-\tau_{MS}(m_1 + m_2)\leq
\tau_{\rm orbital}(P_{\rm init} \rightarrow P_{\rm contact})\leq 
\tau_{\rm cluster}.
\end{array}
\right.
\end{equation}
They define a bound region $\Omega(m_1,m_2,P_{\rm init})$ as a function of 
cluster age. Systems within the allowed bounds are
primordial binaries which would eventually decay, coalesce and form a merger 
in less than the given age but 
which also should still be on the main sequence. A slice of the three dimensional 
region along the plane $m_2 = 0.5$ for a 10 Gyr old cluster is shown 
on Figure 3. The shaded region is the intersection of this plane and 
the region defined by (8).

To find the number of mergers we can then integrate the distribution of primordial 
binaries over the region in three dimensional parameter 
space $(P,m_1,m_2)$ defined by (8);
$$
N_{\rm mergers}(\tau_{\rm cluster})= \int_{\Omega (\tau_{\rm cluster})} 
\xi_1(m_1) \xi_2(m_2) f(P) dm_1 dm_2 dP
$$ 
We normalize it to the number of stars in the two brightest magnitudes 
of MS for a given age and compare to the observations 
of \citet{al95} or to the number of HB stars to compare our predictions
with the fraction of BSs in globular clusters.

\section{Observational data}
In this section we summarize the observational data that we use.
Open cluster data is considered in section 4.1, data on BSs in 
globular clusters and blue metal poor (BMP) field stars in the field are described
in section 4.2, and subturnoff mergers in section 4.3.

\subsection{Open clusters}
Our goal is to see how different ingredients of the model will affect the 
predicted number of single BSs and sub turn-off mergers. To confront the 
predictions of our models with observations we have to bring both to 
the same denominator. From a theoretical perspective, we should use a selection 
function for defining BSs among mergers which closely resembles the one used to select 
observed BS candidates. From the observational side, only single BSs which are
higly probable members of open clusters should be included as BS candidates.

For open clusters we use the catalog compiled by \citet{al95}.
This is the largest compilation of information on BSs in galactic open clusters to date,
although the definition of BSs used is unusual and many candidates lack important
information about binarity and cluster membership, reducing the quality of their database. 
They gathered information on 959 Blue Stragglers candidates in 390 open clusters 
of different ages. A Blue Straggler candidate in their paper is defined as a 
star which appears in a specified region of the H-R diagram (illustrated in their Figure 1). 
Roughly this area is bounded by the ZAMS from the left, the turn-off color from 
the right, and the lowest point on the ZAMS that appears differentiated 
from the sequence of cluster stars from the bottom. The number of BS candidates 
is quantified as a ratio of the number of candidates to the number of main 
sequence stars in the two brightest magnitudes of the upper main sequence.

We need to understand the effects of observational selection both
to estimate the level of contamination of the observed sample and to develop
a theoretical selection criterion that closely resembles the observed sample
for further comparison. We define a clean blue straggler 
sample as all single main sequence stars that are members of a cluster 
under consideration, and are in the specified area of H-R diagram. 
If a clean sample is defined this way, several other types of stars 
may appear as blue straggler candidates and contaminate the sample.

The main contamination source of this is the presence of non-members.
The majority of blue straggler candidates selected by 
\citet{al95} are not confirmed to be cluster members: 567 
stars have one indicator of cluster membership (radial velocity 
or proper motion) and only 161 stars have both indicators of membership. 
Background contamination will be most severe for older systems 
(where the turnoff of the cluster is in a more populated portion of 
the HR diagram) and in systems close to the Galactic plane.
In addition, there is no spectroscopic study of the selected BS
candidates in their sample and therefore binaries will contaminate the sample.
This is particularly important in young clusters. Finally, either
photometric errors or gaps in the CMD for intermediate-aged
clusters can complicate the identification of candidates near the 
turnoff. Such would be the case of Pleiades, considered below.

To get some impression of how the effects described above affect BS 
selection we show CMDs for three clusters in Figure 4. These clusters 
were chosen to represent young, middle-age and old open clusters with 
modern CCD data and large BS candidate populations.

From Figure 4, the claimed BSs in a young open cluster such as NGC 2477 belong
to an extension of the tip of the MS, close to the turn-off point. The main sequence 
is nearly vertical when the most common temperature indicator (B-V) is used.  As a 
result, this area can be occupied either by binaries, or by turn-off MS stars. 
The degree of contamination of the sample in such systems is likely to be high.

An old open cluster such as Berkeley 49 has a 
well defined turn-off, and the area on the CMD occupied by BSs
can not be contaminated by binaries. However, the CMD 
can be severely contaminated by non-members, as is clear from 
Figure 4. Most of the BS candidates in this cluster are probably 
field stars. The degree of field contamination, however, in such systems 
will depend more on the position of the cluster in the sky, rather than its age.

The turn-off point of a middle-age cluster such as M67 is well defined. The contamination
of the BS sample from turn-off stars is not as severe as in young clusters.
In addition, BS candidates lie on the prolongation of MS, which is diagonal for such ages
using B-V as a temperature index. Binaries, therefore, do not contribute 
to contamination as heavily either. However, there is a well-known gap near 
the turnoff that arises from the rapid phase of evolution associated with 
hydrogen core exhaustion. Care must be taken to properly define the turnoff to 
avoid tagging normal stars experiencing core hydrogen exhaustion as blue stragglers.
Although this effect is modest in systems such as M67 (4 Gyr), it becomes more 
pronounced in younger open clusters, and is significant at ages of order 1 Gyr.

To quantify these effects we examined three well studied open clusters
in the \citet{al95} sample: the Pleiades 
(age about 130 Myrs), Praesepe (age about 600 Myr), and M67 (age 4 Gyr). 
For the Pleiades and Praesepe we have extensive membership and binarity studies, 
while the blue straggler population of M67 has been well studied. 

For the Pleiades, \citet{al95} counted 3 BS candidates with 35 stars 
in the 2 brightest magnitudes of the MS. All three stars turned out 
to be binaries. Due to gaps in the upper MS and the
absence of giants it is difficult to determine the turn-off point of such young clusters; 
about half of the Pleiads in the upper MS are binaries. 

For Praesepe they claimed 5 BSs with 30 stars in the 2 brightest magnitudes of the MS. 
Three of these 5 stars appear to be binaries, one is a normal MS star above the 
turnoff gap, and only 1 is a BS. We counted 35 stars in the 2 brightest 
magnitudes of the MS, 17 of which were binaries. 

In M67, where the blue straggler population is the most secure, only a small fraction of BS 
candidates are single stars. This is important, because 
main sequence mergers will produce single blue stragglers, while interactions 
of evolved and main sequence stars will tend to leave white dwarf remnants and 
thus blue stragglers that are still binaries. Hurley et al. (2001) for example, 
who studied BSs in this cluster claim that only 8 of the 28 stars selected from the CMD as BS 
candidates are single stars and/or fast rotators. The rest
are binaries or RS CVn stars (e.g., short period binaries that have not yet merged). 
Two points representing this cluster are shown on Figure 4.  The
filled square represents the actual number of single BSs, while the open one represents the
number of BS candidates for this cluster in \citet{al95}.
As can be seen from the figure, while the open square is consistent with the
general trend for the number of BSs as a function of age, the number of 
single stars (which is relevant for our models) falls close to the 
2-sigma boundary of our prediction.

Unfortunately, it is impossible to do the same exercise on all 
of their clusters, as most of them do not have memberships and/or binarity studies. 
Given the discussion above, we choose to exclude systems younger than 300 Myr from 
our study, on the grounds that detecting blue stragglers in such systems is 
observationally complex and requires followup studies of the binarity of candidates 
that is usually not present. The most believable estimates for a number of BSs are 
obtained for clusters older than 3 Gyr, where the observational definition of the 
turnoff becomes unambiguous. The sample of BSs in many clusters is more prone 
to contamination by non-members, especially in systems close to the Galactic plane,
which makes secure membership studies important. Even in the case of clusters, 
such as M67, in which the turn-off is well defined and 
it is easier to disentangle binaries and TO stars from bona fide blue stragglers, 
care must be taken when comparing main sequence merger models to the data.  The majority 
of blue stragglers are not single stars. Careful spectroscopic study of selected BS 
candidates is thus essential for a rigorous test of the models. We therefore only included 
systems where membership had been confirmed with both radial velocities and proper motions.

\subsection{Globular clusters and Galactic Halo}
There have been numerous studies of BSs in globular clusters. For this paper
we chose to use the \citet{piotto04} catalog. 
It contains about 3000 blue stragglers in 56 globular 
clusters and has been extracted from a homogeneous sample of CMDs 
obtained with the same instrument. The data is given as the normalized 
frequency of blue stragglers to either 
the number of red giant branch or horizontal branch (HB) stars. We used the 
latter in our comparison because the age of HB (and therefore the number of HB stars predicted)
stars is not very sensitive to the assumed age and metallicity for the cluster. 
\citep[]{piotto04} have corrected the BSs and HB star counts for completeness. 

Based on this large sample they concluded that the measured 
frequency of BSs $log(N_{BS}/N_{HB})$ lies between -1 and 0, with a 
couple of outliers at about 0.4. Surprisingly, these the two outliers are also two least luminous
clusters in the sample; NGC 6717 and NGC 6338. As we will see below this high BSs 
population is similar to the fractional population of BMP stars in the 
galactic halo \citep[]{ps00,carn01}. 

There is also a significant anti-correlation between the relative frequency of BSs 
in a cluster and its total absolute luminosity (and therefore mass). The brightest clusters 
are found to harbor the smallest percentage of BSs. However, there is still a large dispersion 
in  $log(N_{BS}/N_{HB})$ at fixed cluster luminosity.

A number of theoretical investigations have proposed that collisions could produce 
a significant number of BSs in globular clusters. However, \citet{piotto04} have found
that there is no statistically significant correlation between $(N_{BS}/N_{HB})$ and the expected
collision rate predicted from the dynamical properties of the clusters. In particular 
post core-collapse clusters that on average have higher central densities behave 
as normal clusters, without clear evidence for a global increase of 
$log(N_{BS}/N_{HB})$.

Field star studies are more challenging and limited in number, but provide 
tantalizing clues about the impact of environment on BSs production.
The population of halo blue metal poor (BMP) stars was studied by
\citet{ps00}. Some stars of their sample and additional
BMP candidates were later investigated by \citet{carn01,spc03,cll05}. 
BMP stars may be distinguished from normal halo 
counterparts in color/color or color/metallicity diagrams. In addition 
to blue colors, BMP stars were found to be different from normal 
halo populations in a few aspects.

There is a very high fraction of binaries with orbital period below 4000 days.
Out of 62 stars in the \citet{ps00} sample, 42 are
probable spectroscopic 1-line binaries, which gives a fraction of $\approx 0.68$. 
Carney at al. (2001) reanalyzed this sample and found 29 spectroscopic binaries 
corresponding to $\epsilon \approx 0.47$. In any case this fraction is much 
larger than that predicted by the \citet{dm91} period distribution
for normal field stars. 

If the total binary fraction of the BMP stars was the typical solar
neighborhood value of 0.5, the fraction of binaries with $P < 4000$ days would be
$\epsilon \approx 0.15$ which is considerably lower than that measured for BMP stars.
Post-MS mass transfer, however, would produce an excess of binaries of BSs 
formed by this mechanism were a significant component of the BMP population. 
This fact was used to argue that BS produced in binaries contributed to BMP sample considerably.

The average eccentricity of binaries in BMP sample is also low. In a small sample
of BMP stars studied by \citet{carn01}, 6 out of 10 stars were binaries with
an average eccentricity $\langle e \rangle=0.11$. This may serve as an additional argument
for the model advocated in that paper, namely one where the majority of the BSs 
in this population form through accretion during the post-MS evolution of the primary.

Stellar abundances can also be used as a test for field BSs candidates. In 
the \citet{carn01} sample 5 out of 6 BMP stars found in binaries were 
Li deficient, and 1 out of 6 was in addition Be
deficient. Both of these elements are fragile and might be depleted in mergers. 
In comparison, 2 out of 3 single BMP stars had normal lithium abundances. 
No data were presented on the lithium abundance of the last star, which was found 
to be in a very long period binary. This suggests that mixing is important for binary 
BMP stars, but less so for the single stars.

We now proceed with the argument made by \citet{ps00} and calculate
the fraction of BSs in the halo relative to the number of HB stars.
BMP stars are speculated by the authors to be a mix of BSs and stars accreted from
Galactic satellites such as the Carina dwarf spheroidal galaxy.
Let $n_{BMP}$, $n_{BS}$, and $n_{A}$ be space density of BMP stars, BSs, and
accreted stars in the BMP sample respectively. Let also $\epsilon_{BMP}$,
$\epsilon_{BS}$ and $\epsilon_{BS}$ are the binary fractions of those sub-samples.
Then we have 2 equations; one for the total number of BMP stars and another for the number
of binaries;
$$
\left\{
\begin{array}{l}
n_{BMP} = n_{BS} + n_{A}\\
\epsilon_{BMP} n_{BMP} = \epsilon_{BS} n_{BS} + \epsilon_{A}n_{A}\\
\end{array}
\right.
$$
This system is resolved to\\
$$
\frac{n_{BS}}{n_{BMP}}=\frac{(\epsilon_{BMP}-\epsilon_{A})}{(\epsilon_{BS}-\epsilon_{A})}
$$
We now have one measured quantity $\epsilon_{BMP}$ and two assumed ones $\epsilon_{BS}$
and $\epsilon_{A}$ to derive the fraction of BSs in the BMP sample, and must also normalize them to
the number of horizontal branch stars.

The uncertainty for the measured quantity is large (compare 
the \citet{ps00} estimate $\epsilon_{BMP} \approx 0.68$ with the 
value $\epsilon_{BMP} \approx 0.47$ if \citet{carn01} 
are correct). \citet{ps00} used $\epsilon_{BS} = 0.87$ and
$\epsilon_{A}=0.15$ which is consistent with the fraction of F and G disk binaries
with orbital periods less than 4000 days. While the latter seems to be a reasonable
estimate, the former needs refinement. The well-studied cluster M67 serves us as 
a useful reference point once again. \citet{hurley01} had 28 BS candidates. If we exclude
the highly eccentric binary, a binary with a period greater than 4000 days, a triple system
and two subgiants, we are left with 23 systems.
Eight of them are single stars, 8 are spectroscopic binaries, and 7 are possibly
RS CVn systems (a close binary with one star being a subgiant). Thus, we can put an upper
limit on the BSs binary fraction to be $\epsilon_{BS} = 15/23 \approx 0.65$.
Given these numbers, we recalculate the expected number of BS in the halo population
relative to HB stars. The results are presented in Table 2.

We expect a comparable number of BSs formed through close binary mergers 
as by post-MS mass transfer events in low density environments (section 5). 
Given this possibility we need to account for possible single BSs in the BMP sample.
We, therefore, include two populations of BSs: single stars and the ones found in binaries. 
We apply a binary fraction of $0$ to the first type, and a binary 
fraction of $1$ to the second type.
The system of equations then becomes;
$$
\left\{
\begin{array}{l}
n_{BMP} = n_{BS,single} + n_{BS,binary} + n_{A}\\
\epsilon_{BMP} n_{BMP} = n_{BS,binary} + \epsilon_{A}n_{A}\\
\end{array}
\right.
$$
This system now has the family of solutions;
\begin{equation}
\frac{n_{BS}}{n_{BMP}}=\frac{(\epsilon_{BMP}-\epsilon_{A})}
{\left ( \frac{\mu}{\mu+1}-\epsilon_{A}\right)}
\end{equation}
that depends on $\mu=(n_{BS,binary}/n_{BS,single})$.

Two families of solutions for two different assumed values of 
$\epsilon_{BMP}$ are shown on Figure 14. The two horizontal lines denote 
the fraction of binaries among BSs in cluster M67. The lower 
line corresponds to $\epsilon_{BS}=0.87$ used by \citet{ps00}
and the upper line to our estimate of  $\epsilon_{BS}=0.65$. The intersection of this
line with the family of solutions equation (9) breaks the degeneracy between the fraction
of BSs in the sample of BMP stars and the fraction of single stars in population of BSs. We
can see from Figure 14 that the minimum fraction of BSs in the BMP sample must be about 40\%
to explain the number of binaries in the BMP sample. The intersection of upper 
horizontal line with dotted line (estimated BMP binary fraction of
$0.5$) gives us the solution that we consider realistic. It would 
imply that 73\% of the BMP population are BSs with a binary fraction of 70\% for the BSs.

\subsection{Subturnoff Mergers}
If the sum of the masses in a binary merger is below the turnoff mass, then 
the product will not be a blue straggler; such objects are more difficult to 
detect as a result. There are two potential observational diagnostics of such 
objects.  Because of the strong mass-luminosity relationship a sub-turnoff merger 
will be significantly less chemically evolved than a single star of the same 
mass and age. In practice this is a difficult test to apply because the 
magnitude of the difference is a function of the initial mass ratio and the 
epoch where the merger occurred, which makes quantitative predictions of the 
number of detectable systems difficult. In addition, distinguishing such stars 
from nonmembers is not trivial.

The surface abundances of merger products may also differ from those of their 
unmerged counterparts, and this does provide a potential test of the population 
of such objects. The potentially observable species most likely to be 
affected are the light elements LiBeB, which are destroyed by proton capture 
reactions at relatively low stellar interiors temperatures (approximately 2.5, 
3.5, and 5 million K respectively). Lithium abundances in stars 
are particularly important because lithium is a 
sensitive diagnostic of stellar mixing and the lithium abundance is relatively 
easy to measure. Severe lithium depletion is the norm in most blue stragglers 
\citep{hm91}but not in all candidates \citep{ryan01,carn01,cll05}.
\citet{ryan02} also found that the In addition, the abundance of $^7$Li in old metal-poor 
stars can be used to test the theory of Big Bang nucleosynthesis 
\citep[e.g.][]{bs85}. The relative surface abundances 
of $^{12}$C, $^{13}$C, and N could also be affected if matter in CN cycle 
nuclear equilibrium is mixed to the surface.
  
There is a substantial body of literature on stellar lithium depletion 
\citep[see for example][]{pinsonneault97}; here we summarize some of the 
main features relevant to using lithium depletion as a diagnostic of mergers. 
Lithium is destroyed in the outer convection zone of open cluster stars with 
effective temperatures below 5000K \citep{iben65}, and there is an efficient 
mixing process that destroys LiBeB in mid-F stars with effective temperature 
between 6250 and 6750K \citep[]{bt86,boesgaard05}. Lithium is usually not 
observable in normal stars much hotter than 7000K 
because it is fully ionized. For open cluster stars with effective 
temperatures below 6200 K there is evidence for a slow mixing process occurring on 
the main sequence that can produce dispersion in lithium abundances among 
open cluster stars of the same effective temperature
\citep[e.g.][]{jones99}. 

The observational situation for halo stars is less chaotic. There is a 
well-defined lithium plateau originally discovered by \citet{2spites82}. 
There is a lively debate in the literature about the possible existence of a 
small dispersion in lithium abundances among halo stars 
\citep{thorburn94,rnb99,pwsn99};
however, any such dispersion is clearly much smaller than the one observed 
in the open cluster case. However, there is a small minority of stars that 
have upper limits well below the typical values for halo stars; 4\% of the 
\citet{thorburn94} hot halo star sample, for example, had upper limits more 
than an order of magnitude below the plateau. If a less stringent criterion 
of a factor of 5 is adopted, the fraction of highly lithium depleted stars 
is approximately 7\% \citep{ryan01,ryan02}. \citet{ryan02} found 
that 3 of the 4 highly lithium depleted stars that they observed had 
detectable rotation, in marked contrast to typical field stars. 
Although this result is strongly suggestive of an unusual history for 
these objects, some care must be taken in interpreting this as direct 
evidence that all highly lithium-depleted stars below the turnoff must 
be objects that have experienced mass transfer. The three rapid rotators 
lie very close to the field turnoff for their metallicity, and may in fact 
be bona fide blue stragglers. These three systems are also binaries with 
orbits consistent with mass transfer from an evolved companion.

Given the sheer complexity of the observational database, we therefore do 
not believe that lithium alone can serve as a diagnostic of the population 
of stellar mergers.  A theoretical calculation of the frequency of such 
mergers, however, could aid in understanding the observed pattern; we 
therefore present calculations for the sub-turnoff merger fraction in section 5.

\section{Results}
Both MS and post-MS binary interactions can lead
to mass transfer or even mergers and in the case
when the final mass is greater than the TO mass of a cluster, 
a BSs can be produced. We begin by estimating the maximum number of mass 
exchanges that could be produced from mergers on the main sequence or 
mass transfer on the RGB or AGB.  We find that the three channels have 
comparable efficiencies; a variety of factors combine to yield total 
populations significantly lower than the upper limits set by this process.

Binaries with small initial orbital period (below approximately 5 days)
and sufficiently low masses (at least one of the components smaller than 1.2
$M_\odot$) may spiral in to form a contact system in less than a Hubble time 
and eventually merge to form a single main sequence star. Although the orbital 
period phase space for main sequence mergers is relatively small, even stars 
well below the turnoff can merge; this increases their number relative to 
interactions that occur after the turnoff.

If binaries are sufficiently wide ($P \ge 5$ days) the timescale
for orbital decay becomes too large for a main sequence merger.  In this case 
mass transfer can occur only when the primary leaves the main sequence. If the 
separation between stars is relatively small ($a \le$ 0.5 AU)
the primary will overfill its Roche lobe and accretion onto the secondary
will occur on the red giant branch. Binaries which are even wider 
(0.5 AU $\le a \le$ 3 AU), can also go through a similar accretion process 
when the primary is on the AGB (the red giant is not sufficiently large to 
overfill its Roche Lobe), or they can accrete a small amount of the 
substantial mass lost in the AGB wind phase. Mass transfer in such 
wider binaries has been studied by a number of investigators 
\citep[e.g.][]{dekool90,il93}. The picture of BMP stars 
provided by \citet{ps00} and \citet{carn01} would be consistent with 
accretion happening during the AGB phase of the primary. In this case 
we would expect BMP stars to be accompanied by CO white dwarfs which have 
mass 0.55 - 0.65$M_\odot$. Mass transfer in post-main sequence systems 
therefore involves a much wider range of orbital phase space than main 
sequence mergers, but a much more restricted range of primary masses.

Given the initial period distribution by the empirical law of
\citet{dm91}, the fraction of binaries with periods
shorter than 5 days is 11\%. The fraction of binaries that
may experience mass transfer on the RGB (with assumed limiting
separation of 0.5 AU) is 8.4\% given the same initial distribution.
Binaries with initial separation between about 0.5 AU and 3 AU will interact when
the primary is on the AGB; their fraction is about 11\%.

For a quantitative upper bound for these channels we chose an age of 4 Gyr,
appropriate for an old open cluster. We estimate that about 18.4\% 
of all binaries would have had a primary more massive than the
current turn-off mass for a MS IMF. Conversely, about 81.6\% had primaries less 
massive than the turn-off mass, and can potentially interact while both 
stars are on the main sequence. However, only about 7.6\% of these 
systems will have a total binary mass higher than the turn-off mass.

The total fraction of binaries which experience mass transfer during the AGB phase of the primary
is therefore about 2\% of all binaries, while the total fraction of the binaries that
may interact while both stars are on the main sequence and produce something more
massive than the turn-off is about 0.8\%. This order of magnitude calculation 
shows that both channels have comparable importance and may contribute to the production of BSs.
It is worth noting that this would be an upper limit for both mechanisms. 
The finite lifetime of blue stragglers will reduce the number of BSs that survive to be 
observed at any given time. Accounting for finite lifetime effects will 
reduce the observed BSs population from both channels. In addition there 
are dynamical effects that could suppress post-MS mass transfer relative 
to MS mergers in some environments. Not all MS binaries that could merge 
will do so because of inefficient angular momentum loss for low mass binaries.
The merger of the secondary and the core of the primary may not be possible if 
the initial binding energy is too low to expel the envelope of a giant, 
which can be important for post-MS interactions. We now discuss each of these effects in turn.

In cluster environment with a moderate collision rate, systems likely to interact 
during the post-MS are much wider binaries that have larger cross sections for dynamical 
interactions than systems likely to merge on the MS. So to a first 
approximation we may expect the reduction of efficiency of BS formation by 
post-MS interactions relative to the MS mergers in such 
environments. The rate of MS mergers could even get amplified 
because, as a result of dynamical interactions, tight binaries get 
tighter. Therefore more binaries with period less than 5 days may be produced, 
and such binaries could then merge on the MS.

For main sequence mergers the orbital decay time is a strong function of 
mass with modern angular momentum loss rates, in the sense that lower mass stars 
experience substantially lower torques than would be 
expected in a simple \citet{skumanich72} braking law. For example, 
from Figure 2 the orbital period of a binary with 
primary and secondary masses of 0.6$M_\odot$ and an initial period of 4 days will 
decay to only 3.5 days in 10 Gyrs. Such a system would not reach contact, 
coalesce and create a merger in a Hubble time. In general, orbital 
decay for binaries with masses lower than $\approx 0.6 M_\odot$ is very inefficient.

Binaries interact in the post-main sequence phase may form common envelope systems 
rather than simply transferring mass; furthermore, not all of the mass of the primary 
star will be donated to the secondary. A common-envelope phase occurs if the accretion 
is unstable (e.g., the Roche lobe shrinks faster than the star losing mass does); 
possible sources of instability are discussed in detail by \citet{hw87}. 
The outcome of this phase may therefore \citep[see][]{dkr93} 
also be a secondary with a mass higher than its initial value, accompanied 
by a He white dwarf in the case of RGB accretion or a CO white 
dwarf in the case of AGB accretion. Some of these post common-envelope 
systems are the precursors of Cataclysmic Variables.

With these estimates in mind, we now turn to computing the rates of main sequence 
mergers and comparisons of the models to the data. We begin in section 5.1 with 
a summary of the model results and parameter variations. We compare our data to 
open clusters, globular clusters, and subturnoff mergers in section 5.2.

\subsection{Model Properties and Parameter Variations}
There are a substantial number of ingredients that enter even into the restricted 
problem that we are examining (main sequence mergers). Furthermore, many of the 
uncertainties are systematic in nature. We therefore begin with an analysis of 
the sensitivity of our model results to the theoretical selection criterion, and 
then examine the variations caused by changes in the physical model in the predicted 
frequency of blue stragglers. We conclude by estimating the resulting theoretical 
errors, which are typically about a factor of two.

We define a reference model using what we view as the most reasonable physical 
choices, and then follow with a discussion of the impact of changes in the
ingredients. For the baseline model we used Miller \& Scalo IMF for primaries and single stars. 
A flat IMF for secondaries was used. The binary fraction was assumed to be 0.5
and a saturated magnetic braking prescription with $m_{crit}=1.2$ and $q_{lock}=0.2$
was adopted. The product in our reference model was supposed to be mixed 
with ZAMS hydrogen abundance. A merger was counted if it was hotter than 
turn-off point of a cluster. All input parameters and their possible range are 
summarized in Table 1.

\subsubsection {Selection Criteria}
In order to test theoretical models we need to be able to define which mergers 
would be observed as BSs. Blue stragglers in different data sets are present 
under different selection criteria, which lead to different and sometimes 
controversial conclusions about their properties. In theory, the predicted 
number of mergers also depends on what we count as mergers. To understand how 
much this selection affects our results we ran model with three different 
selection rules that are relevant for the observational data sets that  we 
are employing (Figure 6). The underlined model is the same in all panels and 
constitutes our standard case (section 5.1.5).

The dotted line denotes the case when all mergers brighter than the turnoff 
of a cluster are counted. Such a criterion is mathematically precise, but observationally 
problematic; there are numerous effects that can produce spurious candidates 
close to the turnoff (see section 3).  The dashed line represents the case 
when we require that a merger product should be at least 0.5 magnitudes brighter 
that turn-off to be counted as a BS. This is similar to the selection criteria 
used in globular cluster studies \citep[for example]{fer03}. 
The solid line is the case when a merger product is defined as any 
object that is hotter than the turn-off and brighter 
than one magnitude below the turn-off. This is chose to reproduce the selection used by 
\citet{al95} and is used to compare our models with their open cluster data. 
It will be shown that differences provided by the changes in the input theoretical 
parameters of the model are considerably milder than the changes in the predicted
number of BSs associated with the selection criteria itself.

\subsubsection{Uncertainties in the Merger Process}
It is straightforward to compute the timescale for binaries to reach contact given 
the binary masses, initial periods, and angular momentum loss rates. However, the 
outcome of the merger process is significantly more difficult to predict. One 
significant assumption, namely that the timescale of mergers is short compared to 
the others in the problem, is held fixed in all of our models.  Given the complexity 
of the problem, we have chosen to study limiting cases about the degree of mixing in 
the final product. 

In one class of models we assume that the product is fully mixed; 
the lifetime of the product will then depend on the degree of nuclear processing in 
the two components prior to the merger, but the blue straggler will begin anew with a 
high central hydrogen abundance. We refer to these as mixed models, and our limiting 
cases are unevolved (zero fuel assumed to have been consumed) and evolved (hydrogen 
consumption prior to the merger being accounted for). Because the luminosity depends 
strongly on the helium abundance, however, the lifetime and luminosity of the merger 
will be a strong function of the prior history of the binary stars. 

In the other class of models one star is accreted onto the other; the merger 
product will then begin its life with a large fraction of its fuel already 
burnt and will have a shorter lifetime. We refer to these as unmixed models. 
Our results are presented in Figure 6. The upper panel shows the 
distribution of hydrogen abundance among fully mixed mergers
created at 1 and 8 Gyr. As can be seen from the histogram, most merger products will 
have hydrogen close to the unevolved value, with a tail extending approximately to $X=0.66$. 
This simply reflects the strong main sequence mass-luminosity relationship. The tail 
corresponds to the limit of two turnoff stars nearing hydrogen core exhaustion.

We use a starting hydrogen abundance of $X=0.66$ for our limiting case; for this purpose 
we ran a set of helium-rich models and calculated the main sequence lifetime as a function 
of initial hydrogen abundance. We then accounted for the higher luminosity and shorter 
lifetime of the product in our measurement of the predicted number of blue stragglers; the 
results are compared with our standard model in the bottom panel. If all merger products 
were fully evolved, we would expect small changes in the number of merger products for 
young systems and a reduction in their number of 0.25 dex at the oldest ages. 

The histograms at the top, however, indicate that the true distribution is peaked at the 
higher (unevolved) level, and that only a minority of merger products would be expected to 
be evolved. As a result, the actual reduction in the predicted number of blue stragglers 
would be well below 0.25 dex, even in in older populations.

The difference between models with mixed and unmixed merged products is illustrated in 
the middle panel. Adopting the lower lifetime of unmixed models instead of the mixed 
models used in our standard model would lower the expected blue straggler fraction by 
about 0.2 dex, with a weak age dependence.

\subsubsection{Binary Fraction, Orbital Locking, and Mass Functions}
For our standard model, we have assumed a Miller-Scalo IMF for the primary 
stars in close binaries and a flat IMF for the relative primary and 
second masses. The impact of changes of the binary fraction and the orbital locking
parameter are illustrated in Figure 8, while the impact of changes in the IMF 
are illustrated in Figure 9. 

From Figure 8, the effect of change of the binary fraction or orbital locking
parameter within reasonable limits is rather mild. Changes in the assumed IMF for 
primaries have only a modest effect (0.16 dex deviation from the standard 
case at most) because the number of blue stragglers is compared to the 
number of stars on the upper main sequence, and the 
majority of mergers that are counted as BSs occur for primary masses close to 
the turnoff. In the case of a flat mass function for the secondary, the Salpeter 
IMF predicts more mergers, because the effect of the decrease in the number of MS stars in 
the upper main sequence (to which we normalize our results) is larger than the 
decrease in the number of binaries with efficient angular momentum loss (which 
extends further down the main sequence).

The choice of the distribution of secondary masses relative to the distribution of primary
masses has a larger impact on the models, up to about 0.3 dex in young clusters;
with much smaller effect (order of 0.1 dex) in old populations. Adoption of a 
Salpeter IMF suppressed the merger rate for two reasons. The secondaries are lower 
in mass and are less likely to donate sufficient mass to make a blue straggler. 
In addition, lower mass secondaries experience less efficient angular momentum loss, 
and the binary systems that contain them will have narrower range of initial periods
of binaries that will merge. 

\subsubsection{Magnetic braking}
The impact of choosing different magnetic braking prescriptions is shown in Figure 10. 
Saturation effects limit the rate of angular momentum loss in our standard model, 
resulting in a lack of mergers for young systems. The more efficient unsaturated braking 
mechanism leads to much more efficient production of mergers early on 
(at the factor of 5 level), and a corresponding decrease of up to 0.4 
dex at older ages (because the systems in question have already merged and 
the product has left the main sequence). 

We also incorporate the known absence of a magnetized wind for upper main sequence stars 
in our model; the exact mass threshold, however, is somewhat difficult to measure precisely. 
From Figure 10, an increase of the limiting mass at which stars experience braking from 1.2 
to 1.4 solar masses increases the number of binaries which can spiral in towards contact, and 
also contributes to the production of mergers for systems with turnoff masses greater than 
1.2 solar masses. This has little impact on the oldest systems, but could increase the 
predicted number of blue stragglers by up to 0.4 dex in the youngest systems. As 
we discuss further in section 5.2, the current data is much more 
consistent with a saturated braking law than with unsaturated, 
when only single BSs are considered. 

\subsubsection {Standard model and theoretical error analysis}
To calculate the theoretical uncertainties we compared 
the models described in the previous section with our reference model 
and calculated the quadratic deviation. The parameters considered were the 
primary IMF, fraction of binaries, 
maximum mass of a MS star that experiences magnetic braking, type of merged 
product (mixed or unmixed) and the minimum mass ratio at which the 
secondary is tidally synchronized.

For all models a flat mass function for the secondary was used and we adopted a theoretical 
selection that resembles the ones used in \citet{al95}.
We did not include evolutionary status  within mixed models for two reasons. 
First, the effect of this parameter is relatively small even for most evolved 
mergers; second, the number of very evolved mergers is
not high as shown on the Figure 6. We decided to disregard unsaturated braking in calculating
uncertainties, because this model is grossly inconsistent with spindown studies of single stars.
From \citet{terndrup00}, the uncertainties in the empirical braking law are small enough
(20\% at the 1$\sigma$ level) that they should not affect our results. 
We did not include model in which secondaries are represented by the same mass function 
as primaries, because we think that such an assumption for close binaries 
which contribute to merger production (with periods less than 5 days) is physically unmotivated.

In cases where the difference was systematic, we assumed that the difference 
between the models corresponded to an effective $2\sigma$ error. In our models this applies to 
the cases of the critical threshold for angular momentum loss 
$q_{lock}$ and mixed compared with unmixed models. We treated binary fractions in the 
range $\epsilon = 0.3$ and $\epsilon = 0.7$ as plus and minus $1\sigma$ cases, as 
well as differences in the IMF between our standard MS IMF, the Salpeter IMF, and the 
M35 IMF respectively. 

Our standard model with $2\sigma$ boundaries are illustrated in Figure 10 
together with the data by \citet{al95}. This data was corrected
for the presence of BSs in binaries, and only highly probable members of open clusters
were considered, as discussed in the next section.
We also show the data for Praesepe (based on one star!) and M67 for all 
blue stragglers and the single blue stragglers alone. 
Once corrected for binarity and membership, the predictions of our model match
the data very well, as described in more detail in the next section.

\subsection{Comparison of model predictions with data}
\subsubsection{Open clusters}
The total blue straggler fraction in the entire sample of open 
clusters claimed by \citet{al95} is substantially higher than our theoretical estimates. 
However, as discussed previously, the blue straggler
fraction is reduced (and more reliably estimated) in the subset of systems
with good information about membership. The total blue straggler 
fraction of this sample is at the high end of that (but within) the 
$2\sigma$ theoretical error range. However, there is no a priori reason to
expect all blue stragglers will be produced from main sequence mergers. 
In particular, blue stragglers produced by post-MS mass transfer would 
still be binaries, while main sequence mergers would produce single blue stragglers.
We should therefore really be comparing the theoretical predictions 
to the number of single blue stragglers. 

In the most comprehensively 
studied cluster, M67, only $8/28$ BSs are single. If we assume that 
this proportion is characteristic of all open 
clusters, then we should reduce the observed BSs fraction by this factor.
In order to compare with the open cluster data we therefore proceed as follows.
We included the data from the \citep{al95} study that had good membership information (from
both proper motions and radial velocities) and assumed that the fraction of single
blue stragglers was the same in all clusters as in M67. We then divided all clusters into 
3 equal in log(age) bins, calculated the total number of single BSs 
and the total number of MS stars in the brightest 2 magnitudes. 

The results are illustrated in Figure 11. The open circles represent the individual 
clusters and the solid points with one $\sigma$ error bars indicate 
the average results for each bin. The solid line is the prediction 
of our base theoretical case, while the dashed lines represent the $2 \sigma$ theoretical
error ranges. The agreement between the number of single blue stragglers observed and 
the theoretical predictions is extremely good. We interpret this 
as evidence that approximately one third of the \it bona fide \rm  blue stragglers 
in open clusters are produced by the main sequence merger of primordial close 
binaries, with an uncertainty of 0.12-0.2 dex at the one $\sigma$ 
level depending on the age of the cluster. In the next 
section we will address the question of whether this ratio depends on stellar environment.

We find this agreement encouraging, and also note that the age trends in the models are
qualitatively reproduced in the data. We can also therefore examine the question of
what constraints can be imposed on the theoretical model ingredients from the open cluster
data set. In order to confidently use the data to discriminate 
between different classes of theoretical models, it would be helpful to determine whether
the binary fraction of BSs in M67 is representative of the open cluster sample as a whole, 
or if it depends on the cluster age or dynamical properties.  In the discussion below we will
proceed with the assumption that these differential effects are not present.

As can be seen from Figures 7-10, the sensitivity of our model to most of the
theoretical ingredients is rather mild. From Figures 7-9, the 
degree of mixing in the merged product, IMF,  and values of $\epsilon$ 
and $q_{lock}$ within reasonable limits does not affect the prediction of our model much.
However, there are some cases where the adoption of different theoretical 
assumptions would significantly degrade the agreement between observations
and theory. In particular, adopting an unsaturated (Skumanich type) braking law (see Figure 10)
both produces too many mergers for clusters younger than 1 Gyr and too few mergers
for clusters older than about 5 Gyrs at a statistically significant level.
There is a straightforward physical explanation for this behavior. Saturated and unsaturated loss laws
differ only at relatively short orbital periods, and over a Hubble time 
sufficiently short period systems would be expected to merge for either.  As a result the total
number of mergers over a Hubble time is comparable, but the timescale for a merger to occur is
longer in the unsaturated case. Unless there is a strongly age-dependent change 
in the binary fraction of BSs, we conclude that the unsaturated braking law 
is severely disfavored by the data. A more comprehensive study 
of the binarity of blue stragglers, especially in older populations,
would permit more stringent tests of the angular momentum loss law in close binaries.

\subsubsection{Globular clusters and halo field stars}
Globular clusters provide another natural test of our models, and the sample of 
\citet{piotto04} is well-suited to provide a basis for comparison. Compared to the 
open cluster case, blue stragglers can be more easily disentangled from contaminants. 
The number of upper MS stars is not a good calibration point for globular clusters, 
however, because of crowding. As a result we adopt a different normalization 
for the theoretical calculations that has been widely used in globular 
cluster studies, namely the blue straggler population 
relative to the number of horizontal branch stars. We define a BS as an object 
merger more than 0.5 magnitudes brighter than the turnoff. This choice 
resembles the observational selection criteria used in globular clusters
\citep[][for example]{fer03}. 

The number of horizontal branch stars relative to turnoff stars is a straightforward 
function of the relative lifetimes of evolved stars in different phases. The lifetime 
of the HB phase is well-constrained because stars arrive on the HB with a narrow range 
of helium core masses.  In our models the number of HB stars which are single is given by;
$$
N_{HB} = \int_{m_{to}(\tau_{cluster} - \tau_{RGB})}^{m_{to}(\tau_{cluster}
-\tau_{HB} - \tau_{RGB})}\xi (m)dm
$$
where $\xi (m)$ is the mass function used for single stars, 
$m_{to}( \tau )$ is the turn-off mass
corresponding to a given age $\tau$, $\tau_{HB}$ is the lifetime of a star on horizontal 
branch, and $\tau_{RGB}=1Gyr$ is taken as the lifetime on the red giant branch. The results that 
follow are not very sensitive to the adopted RGB lifetime. In order to get a sense of 
the impact of age effects we considered models with cluster ages between 8 and 12 Gyr
with solar metallicity.

Intermediate-period binaries will experience interactions on the first ascent RGB, 
and we do not include stars in this category in our estimates of the HB population.  
Wide binaries (with orbital separations exceeding ~ 1 AU) will avoid mass transfer 
and/or common envelope formation, and stars in wide binaries are thus able to 
evolve into HB phase. 

For the lifetime of a star on the horizontal branch, we took results from \citet{ld90}. 
Given the range of masses and composition, this lifetime ranges from 85 Myr to 115 Myr, we 
considered 3 models with $\tau_{HB}$ equal to 85 Myr, 100 Myr, 115 Myr.

The results are shown in Figure 11. Three curves denote models with different 
assumed HB lifetimes. The shaded region approximately corresponds to the 
frequency of BSs in globular clusters \citep{piotto04}. The prediction 
of our model traces the upper boundary of the shaded region. No strong 
correlation with age is expected in this age range, and there is no strong observational 
evidence for age-related variations in the blue straggler populations of globular clusters.
Our predictions trace the upper boundary of BS frequency in globular clusters.
This result is in striking contrast with the frequency of BMP stars in 
the galactic halo. The excess of observed BMP stars to predicted single blue 
stragglers is a factor of 4. Significantly, when the sample is restricted 
to exclude known blue straggler binaries the agreement between theory 
and observations is dramatically improved; the single blue halo 
field objects and the single M67 blue stragglers are a factor of 1.7 and about 1.3 greater 
than our predictions, consistent within the theoretical errors.

A synthesis of the observational data therefore permits the following working hypothesis. 
In sparse environments such as open clusters or the halo field, approximately 1/3 
of the blue stragglers arise from main sequence mergers. The remaining systems come from 
mass transfer involving RGB and AGB stars. In the dense environments of globular clusters 
wide binaries are destroyed, and the predicted number of blue stragglers from main sequence 
mergers is consistent with the measured number of blue stragglers. In this picture we rely 
heavily on using the binary status of blue stragglers as a proxy for their channel of origin; 
single stars are ascribed to main sequence mergers while binary stars are ascribed to mass 
transfer on the RGB or AGB. There are two general predictions from such models.  Blue 
stragglers in globular clusters should be primarily single stars, and the majority of 
claimed open cluster BS candidates in the \citet{al95} sample are either 
background contaminants or binary stars.

The preceding comment does not imply that dynamical production of blue stragglers in 
globular clusters does not occur. The peculiar bimodal radial distribution of blue 
stragglers in M3 \citep{guh94} and 47 Tuc \citep{fer04} provide clear evidence 
that there are two distinct mechanisms for formation. In 
the centers some are probably created by collisions, and in the outer layers 
by some other mechanism, which should become predominant when the collision rate is low
\citep{mapelli04}. 

Our investigation demonstrates that our model mechanism for mergers is able to produce a 
sufficient number of BSs even before collisions are accounted for; this implies 
that main sequence mergers of primordial short period binaries must be accounted 
for in studies of blue stragglers in globular clusters. Sophisticated dynamical 
models which take into account production of close binaries, destruction of wide 
binaries, evaporation of low mass stars and mass segregation will provide valuable 
insight into the relative roles of dynamical effects and binary mergers. 

In our view it is likely that the net effect of dynamical processes is to lower the blue 
straggler fraction, and that the fraction of collision products in the BSs
is likely to be modest. This is consistent with dynamical results presented for 47 Tuc,
where approxitely 10\% of the BSs were estimated to have a collisional origin by
\citet{mapelli04}. 

\subsubsection{Under TO mergers, possible connection with lithium depleted stars}
The bulk of our discussion has focused on merger products that can appear as 
blue stragglers. However, a significant fraction of main sequence mergers will 
produce objects that are below the cluster turnoff. We illustrate the 
age dependence of this phenomenon in Figure 12, where the fractional 
population of subturnoff merger products in half-magnitude 
bins is plotted as a function of bolometric magnitude for
ages of 0.6, 4, and 12 Gyr. These can be roughly thought of as being 
representative of intermediate-aged open clusters (e.g. the Hyades), old 
open clusters (M67), and the halo field respectively. The average
population in the 2 magnitudes below the turnoff increases gradually with age, 
from about 1.5 percent for the 0.6 Gyr sample to about 4 percent for the 
halo turnoff sample. In all cases this is less than
the proportion of highly lithium depleted stars, implying that main 
sequence mass transfer cannot solely explain the presence of 
highly lithium depleted stars.

It should be noted that this is an upper limit; we have chosen the mass function 
that maximizes this population. An uncorrelated primary/secondary IMF would 
produce much lower mass merger products; a steeper than flat relative 
IMF would make more blue stragglers and fewer subturnoff mass products.
It is difficult to provide any quantitative estimates of lithium depletion
in the AGB mass transfer scenario and compare it to the amount of depletion
expected in MS mergers. The relative depletion of different light species
(Li,Be,B) that burn at different temperatures should be investigated, as 
it could be used as the test of underlying cause of the anomalous abundances.

We, therefore, conclude that a background population at the few percent 
level should be accounted for in lithium studies, but that it would be desirable 
to establish the presence of a depletion pattern inconsistent with 
mild mixing before removing such objects from abundance study samples.

\section{Summary and discussion}
We have investigated one channel of the formation of single blue stragglers in 
low density stellar systems. It has long been known that the merger of close primordial binaries 
can be a plausible origin of Blue Stragglers. Early works on the subject 
\citep{vilhu82} and \citep{stepien95} demonstrated that even given the 
uncertainties associated with the modeling of such wind, this mechanism 
can explain the observed number of contact systems. Our major results both quantify 
the importance of main sequence mergers and clarify the role of environment in BSs production. 
We have also included calculations of the production rate for subturnoff merger products, 
which are an interesting and surprisingly numerous population.

The comparison of theory to observations is significantly complicated by 
how a BSs is observationally defined.  In fact, we found that the predicted 
BSs fraction was affected more by the mapping of theory to observations 
than by changes in any of the parameters in the theoretical models.  
We conclude that to make quantitative predictions theoretical models 
must use a definition of a BSs as close to that used in observational 
studies as possible.  This explains the differing normalizations 
that we used when studying different populations. It is nonetheless 
possible to define the ratio of predicted to observed systems in 
different environments, and we believe that differential changes 
in these ratios will be less sensitive to observational selection 
effects than the absolute number.   

As it stands now, our model predicts about 1/3 of the 
number of BSs claimed by \citet{al95} for intermediate age clusters (1 - 5 Gyrs).
This number corresponds well to the observed fraction of single blue stragglers
in well studied populations (such as M67) and the inferred fraction of single 
stars in the BMP population
of the Galactic halo. After taking into account only single BSs and correcting
the data by the same binary fraction as observed among population of BSs in M67,
the model reproduces the observed number of BSs in Open Clusters
of single blue stragglers by \citet{al95} quite well. 
The tendency for the BSs fraction to rise for medium aged clusters (0.5 - 3 Gyrs) and flatten 
for older ones (3-10 Gyrs) is seen in both the data and the prediction 
of our model. Data on binarity and membership of BSs in open clusters would permit 
more stringent tests of our model.

Based upon our theoretical error estimates, the overall uncertainty in the 
predicted number of mergers is at the factor of two level. The major 
ingredient not included in this error budget is the binary period distribution. 
If we had used the binary period distribution for old field stars, where short 
period binaries are relatively rare, we would have predicted a much lower 
binary fraction.  However, because we used the period distribution in young 
clusters, the presence of a low period spike in the initial period distribution 
of close binaries 
produces a number of mergers close to the observed number of single 
Blue Stragglers in open clusters.  Our models are consistent with the field 
star data because mergers remove primordial short period binaries. 
As a result the short period spike is absent in the theoretical models 
for older stars.  We believe that this is consistent with the idea that 
the missing systems have undergone mergers.  This conclusion is an 
indirect one, and it would be helpful to test it in other ways. 
The strong mass dependence of the merger timescale that is predicted 
by empirical models of stellar angular momentum loss may provide such 
an alternate test of the merger hypothesis.

In older models of magnetic braking lower mass stars in close binaries 
have comparable torques but much lower angular momenta than higher mass 
binary systems.  As a result, there is a tendency for them to merge over 
comparable, or even shorter, timescales than higher mass binaries. 
However, the open cluster data for single stars requires a rapid 
decrease in the loss rate for decreased stellar mass. This implies 
that the fraction of short period binaries with masses below 0.6 solar 
mass should decline with age much more slowly than the comparable fraction 
for solar mass stars.  If this can be confirmed it would be a strong piece 
of evidence that it is mergers that are responsible for the difference 
between the open cluster and field populations.  
The origin of this bimodal period distribution for binaries 
is a separate and interesting theoretical question, and for constructing 
predictive models of BSs populations its dependence on mass and composition 
should be investigated. 

The predicted number of BSs as a function of age can also be used to test 
magnetic braking prescriptions.  We find that an unsaturated magnetic 
braking prescription is inconsistent with the age trends in the data.
The unsaturated prescription produces too many BSs at young ages; by the time
old ages are reached the precursors have all been used up leading to a deficit
in the number of mergers. When this happens, we should necessarily expect
a decline in the number of expected BSs (as seen in Figure 10). 
The saturated braking law, on the other hand, does predict the correct shape 
and the number of BSs for all 3 bins. We include a magnetized 
wind only for low mass stars; the data suggests that adopting a cutoff
at the lower end of the plausible range ($1.2 M_{\odot}$) is the best 
current choice.  In our view the presence of BSs in young systems (less 
than 300 Myr) is unlikely to be a binary evolution effect; rotationally 
mixed models are more physically motivated than invoking a magnetized 
wind above the break in the Kraft curve.

Our model predicts that MS mergers can account for about 1/4 of 
Blue Metal Poor Population of stars observed 
in Galactic halo, albeit based on limited statistics. According to \citet{ps00} 
and \citet{carn01} the majority of these systems are members of binaries, 
and, therefore, were probably formed as a result of accretion from a giant on 
a main sequence star. If these systems are associated
with BSs, then post-MS mass transfer accounts for about 3/4 of BMP population. 
Such systems should be members of relatively wide (separation 
of about AU) binaries. The rest - 
about 1/4 of BMP stars - should then be single or contact main sequence binaries, 
and are contributed by mechanism we are discussing. Our model looks consistent 
with the number of single BMP stars in galactic halo. However, better statistics
and abundance studies would again be desirable.

Even though we realize that dynamical effects may be important for the
formation of BSs in globular clusters, we 
compared our predictions with GC data by \citep{piotto04}. With appropriate 
selection criteria imposed on MS mergers, we found that they can account for all of the 
BSs, even in the most rich globular clusters. This MS merger rate is less sensitive to 
changes due to dynamical processes (collisions) in globular clusters than 
post-MS mass transfer because close binaries (with periods of 1-5 days) 
involved in MS mergers have a much lower cross sections for encounters than
wide binaries, and statistically get tighter as a result of collisions, possibly even enhancing
MS merger rate. Wide binaries (with periods of years) have larger cross sections 
(more encounters) and statistically get wider, this should decrease if not 
completely eliminate the production of BSs by mass transfer from evolved primaries. 
We find it extremely intriguing that the predictions of our model trace the upper
boundary of number of BSs in globular clusters. 

It looks possible that mergers of close binaries may be the primary source 
BSs in globular clusters. Detailed dynamical models are required 
to assess this possibility.  Dynamical effects would modify the 
period distribution of the close binary population. 
The binaries located in the lower spike of period distribution
will shift to lower periods, and therefore produce mergers early on in the lifetime of a 
cluster. Such mergers would finish their main sequence life, and would not be observed 
at the typical age of a globular cluster. This (together with mass segregation) might 
explain the peculiar radial distribution of BSs in M3 and 47 Tuc by suppressing BSs production at 
intermediate radii. In the outer 
regions where collisions are not so common, our mechanism should predict 
population of BSs relatively well.  In this sense our results are compatible 
with the dynamical study of BSs populations in 47 Tuc by Mapelli et al. (2004). 
They found that only about 10 \% of the BSs in 47 Tuc came from collisions, 
and that the unusual radial distribution seen in that system could be 
explained by dynamical effects if there was a substantial source of BSs 
arising from primordial binaries.  Dynamical studies of this type will 
be useful for understanding why peculiar bimodal distribution were found only in a couple of 
clusters, and how much direct collisions of single stars 
contribute to the production of BSs in the center. Our models are 
complementary, in the sense that we can provide a plausible theoretical 
estimate of the BSs production rate itself.

We found a predicted population of subturnoff 
mergers comparable to the ultra Li depleted population in halo field stars.
MS mergers could therefore be important in high precision abundance studies. 
However, the predicted merger fraction is well below the observed data in open cluster systems.
We believe that the best method for determining the contribution of mergers to MS 
abundance anomalies is to compute the predicted depletion patterns on a 
case-by-case basis and to compare directly with the data.

We want to conclude this paper by listing observations we consider important
for future progress in this field. More 
data on BSs in open clusters would
be especially useful to comparison with our model. In particular, 
the binarity of BSs may be the best observational diagnostic of their origin. 
If the binary fraction can be measured reliably as a function of age it may be 
possible to empirically infer the main sequence merger rate as a function of age, 
and therefore constrain the models much more tightly. 
The correspondence between our results for open clusters and BMP field 
stars is interesting, but it relies heavily on data obtained for a small 
sample.  More generally, an empirical constrain of complete distribution function
$f(m_1,m_2,P)$ in the period range of 1-10 days for binaries in young 
clusters would provide a valuable ingredient for theoretical modeling; 
in effect this would test the assumptions that we have made about the 
universality of this function. In particular, the difference between the 
primary star IMF and the primary to second star mass ratio is important 
for predicting which mergers will produce BSs and which ones will be 
subturnoff mergers.  In addition to its implications for the problem 
that we are considering, such data could potentially constrain theories 
of star formation.

\input{srefs.tex}
\input{pics.tex}

\input{tab1.tex}
\input{tab2.tex}

\end{document}

%% file: pics.tex
\begin{figure}[t]
\plotone{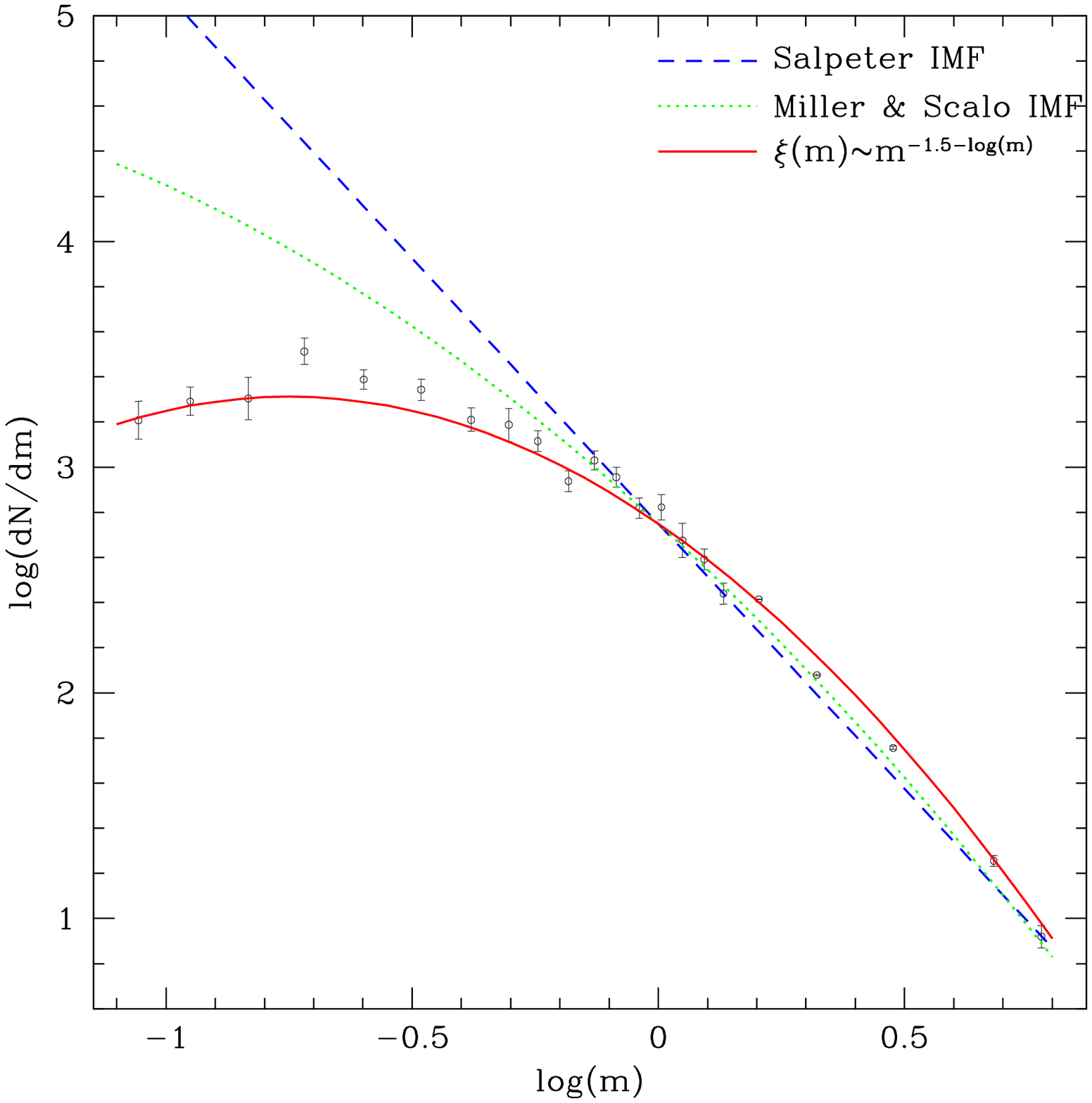}\caption{Mass functions used in the computation of
merger rates. As described in the text, we used the Miller \&
Scalo (1979) function (dotted line) as a base case, but also explored the
effect of using the Salpeter (1955) function (dashed lines) and a third
function which approximates the mass distribution of stars in the open cluster 
M35 (solid line) as presented by Barrado, Stauffer, \& Mart\'{\i}n (2001); the data
for M35 are dhown as points with error bars.}
\end{figure}

\begin{figure}[t]
\plotone{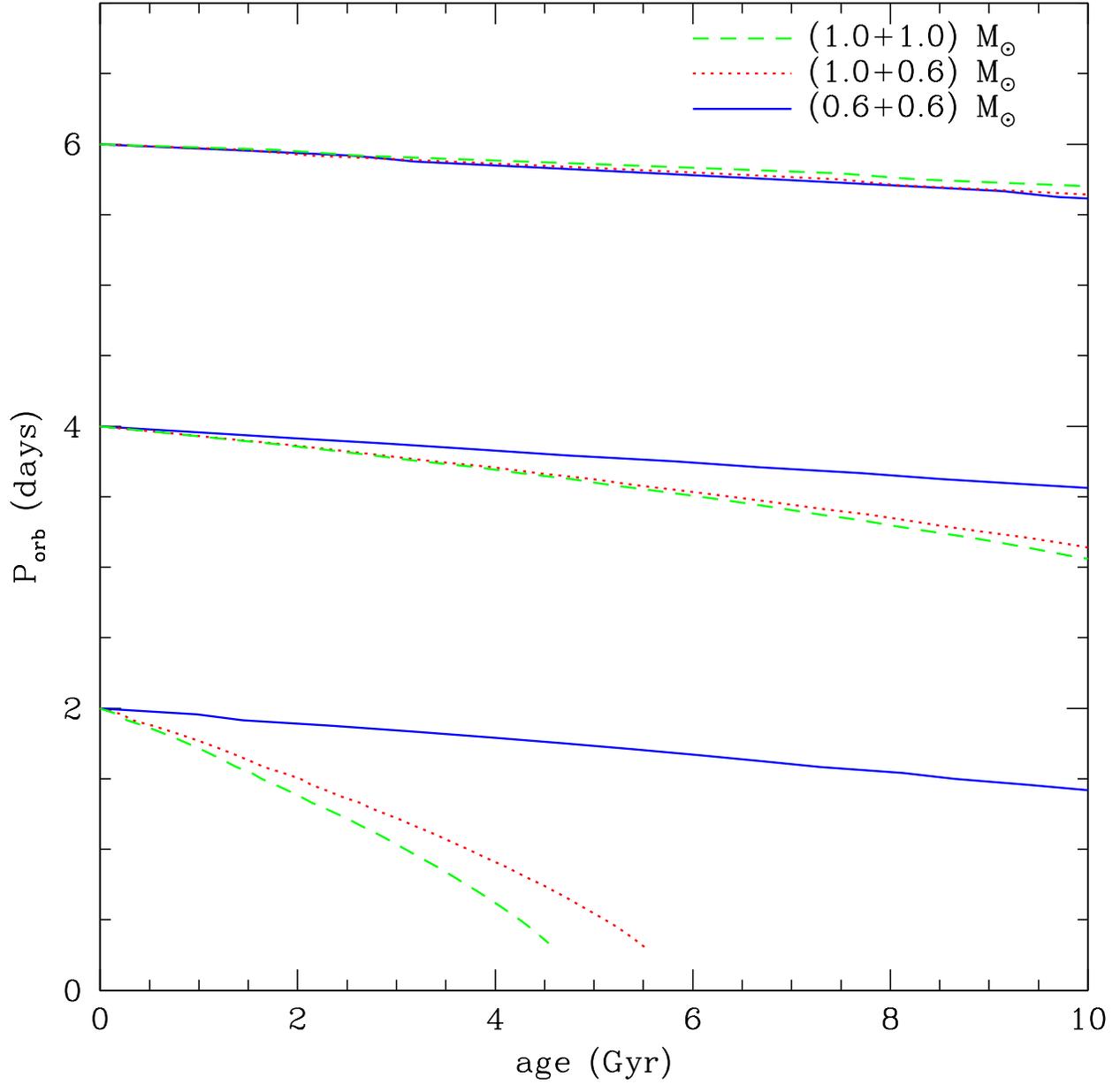}\caption{Orbital period as a function of age for three binaries
with different masses; $(1+1)M_\odot$, $(1+0.6)M_\odot$, $(0.6+0.6)M_\odot$. 
Initial orbital period was chosen to be 2, 4, 6 days.}
\end{figure}

\begin{figure}[t]
\plotone{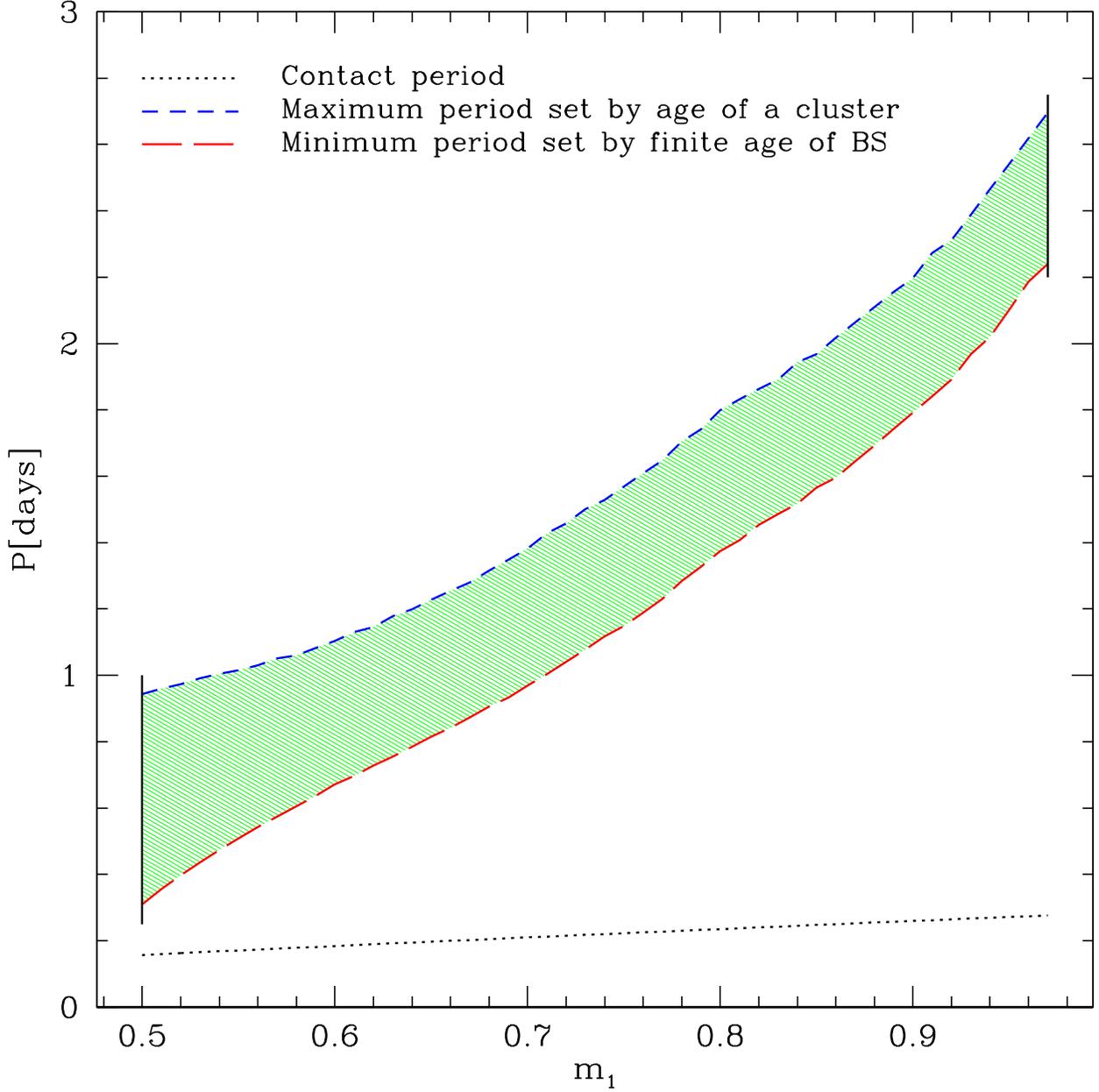}\caption{Intersection of a region in parameter space points of which
correspond to primordial binaries that produce observable BSs for given age of a cluster
and plane $m_2=0.5$. An age of a cluster for this plot was 10 Gyr. 
Different limitations are shown as a function of mass of the primary star. 
If the binary has an initial period below the shaded region, it will spin down too quick, 
and the merger product will end its main sequence lifetime before the time of observation. If,
on the other hand, the initial period was above the shaded region, the binary
will not be able to achieve contact at the age of the cluster. Two
solid vertical lines denote the lower mass limit for the primary (more
massive than secondary) and the upper mass limit (less massive than turn-off
mass for cluster).}
\end{figure}

\begin{figure}[t]
\plotone{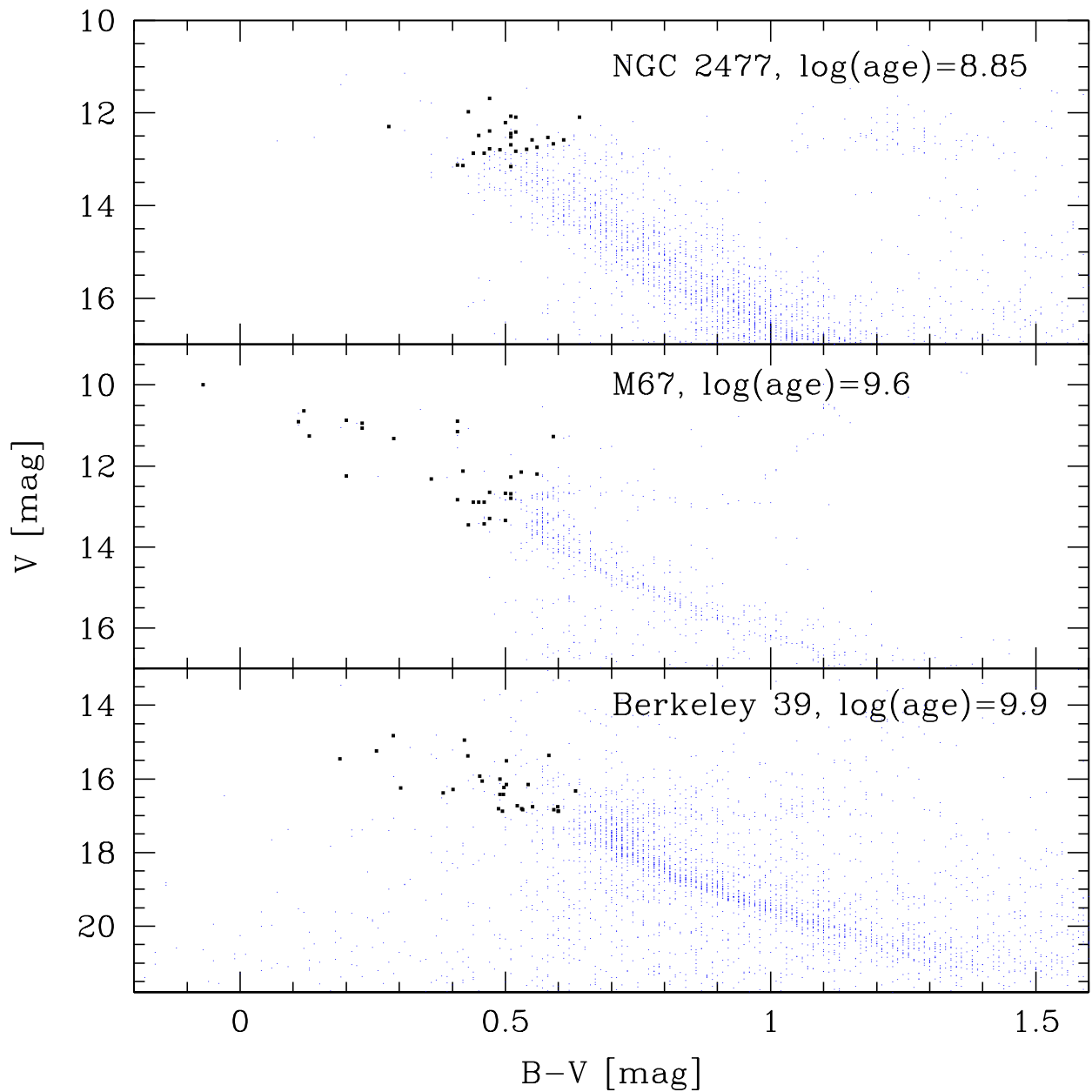}\caption{Upper panel: a CMD of the young open cluster NGC 2477
($log(age)\approx 8.85$) is shown on the top panel. CCD data 
is taken from Kassis, Janes, Friel \& Phelps (1997). The claimed number of BSs is 28.
Middle panel: a CMD of M67 ($log(age)\approx 9.6$). CCD data is taken from 
Montgomery, Marschall \& Janes (1993). The claimed number of BSs is 30. 
The actual number of BSs is 28, but only 8 of them are single stars (see Hurley et al. 2001)
Lower panel: a CMD of the old open cluster Berkeley 39 ($log(age)\approx 9.9$). 
It is claimed to have 25 BSs by Ahumada \& Lapasset(1995). CCD data is taken 
from Kaluzny \& Richtler (1989).}
\end{figure}

\begin{figure}[t]
\plotone{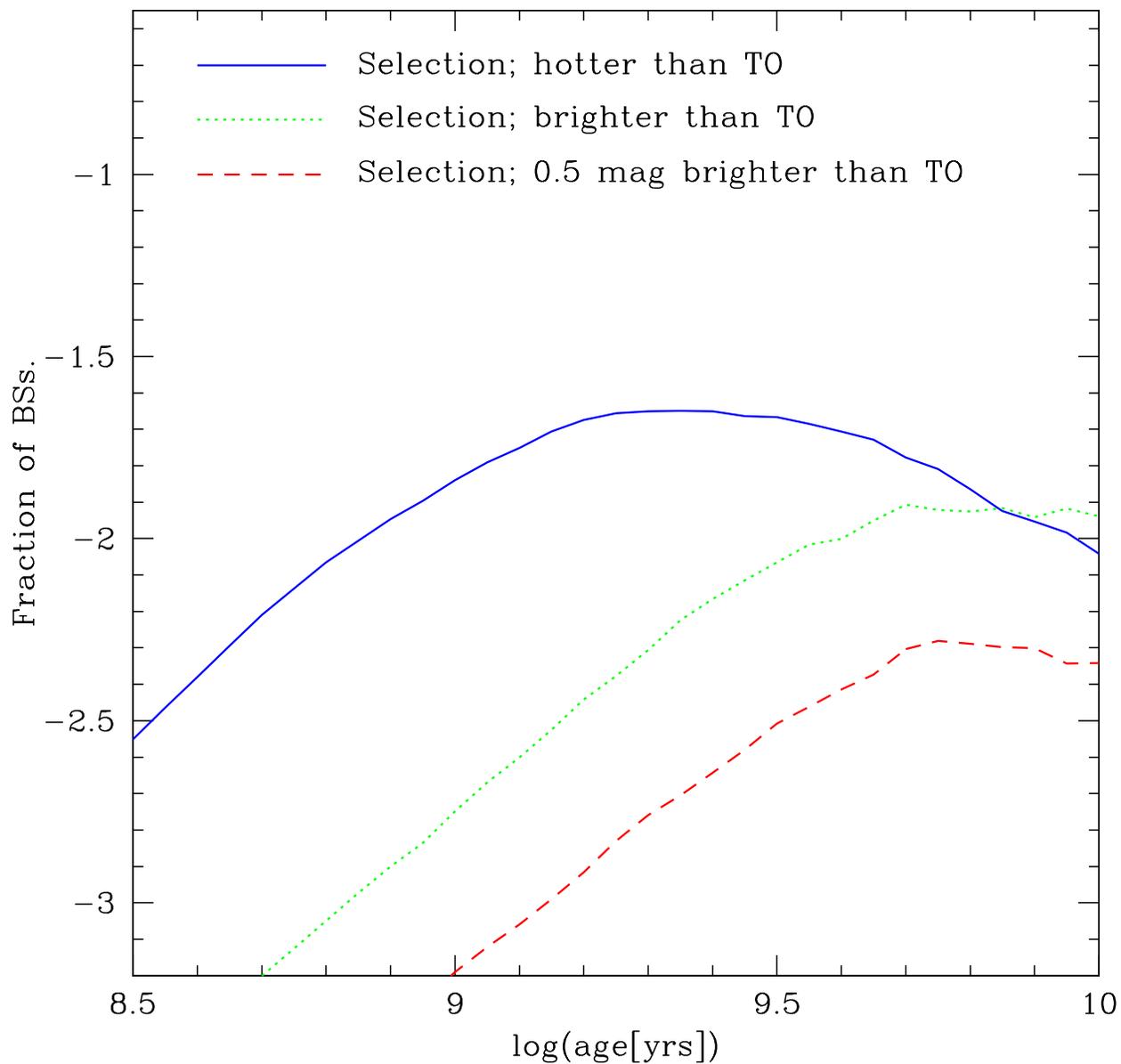}\caption{The predicted number of single blue stragglers when 
different selection rules are adopted.
In all cases, the number of blue stragglers is normalized to the number of stars
in the two brightest magnitudes of MS as function of age. The solid blue line
represents the selection of mergers which are hoter than turn-off.
Dashed and doted line represent selection by luminosity of a merger.
Mergers are required to be brighter than turn-off (dotted line) or brighter 
by at least 0.5 magnitudes than turn-off (dashed line).
}
\end{figure}

\begin{figure}[t]
\plotone{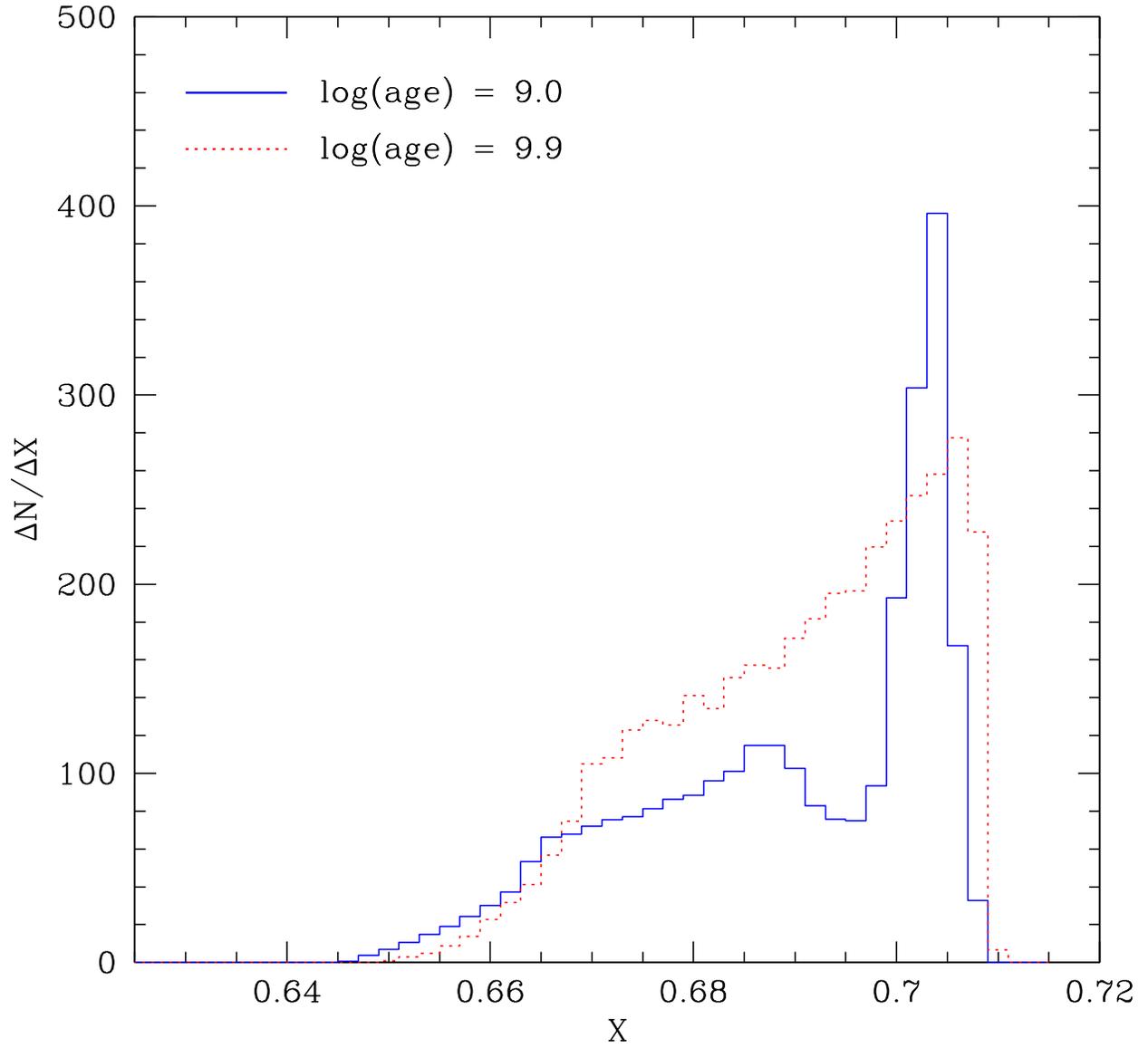}\caption{Distribution of hydrogen abundance among fully mixed mergers for 2 
different
ages of the stellar population. The purpose of this plot is to demonstrate that 0.65-0.66 is about lower
hydrogen abundance of such objects.}
\end{figure}

\begin{figure}[t]
\plotone{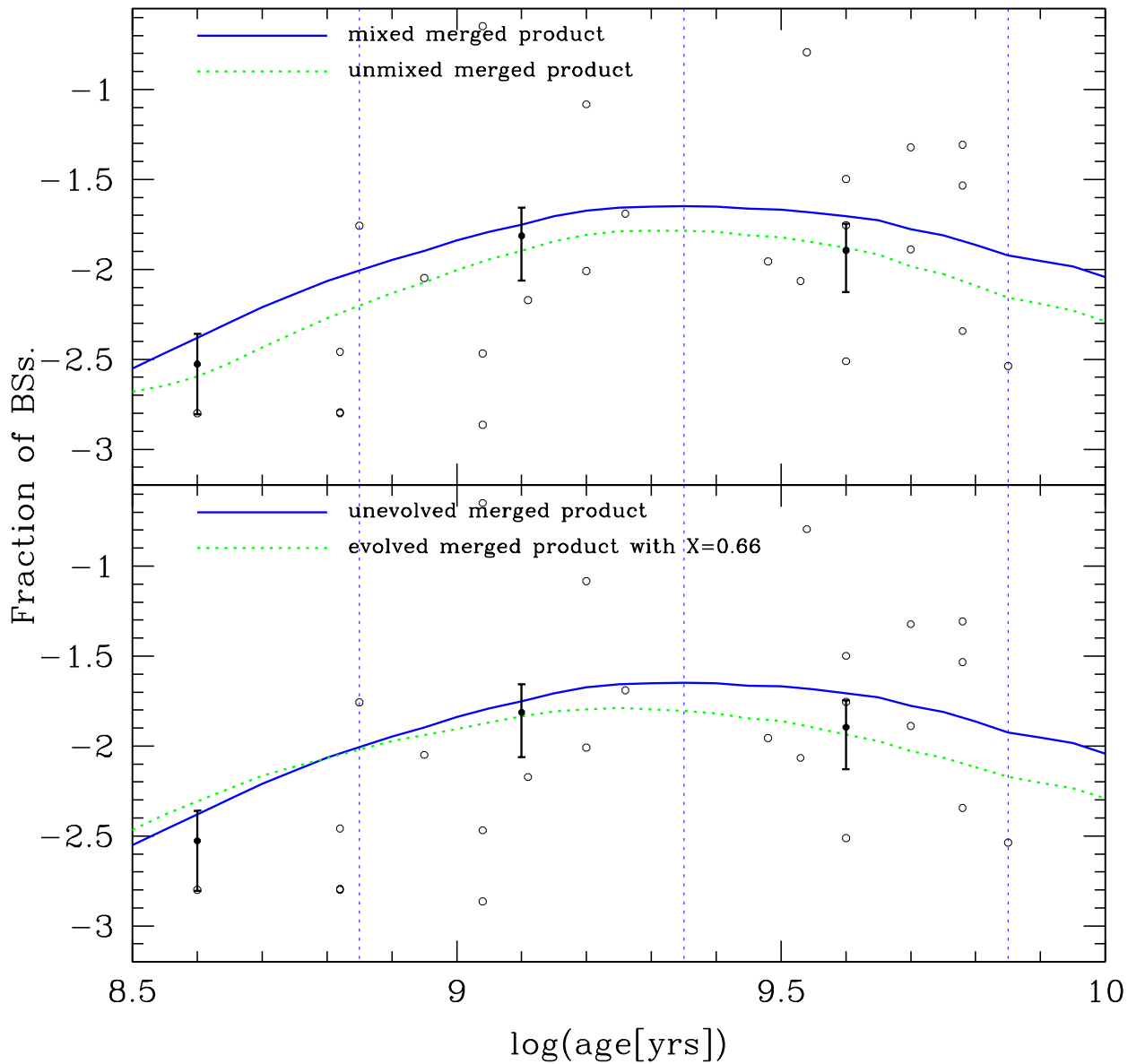}\caption{The predicted number of single blue stragglers normalized to the number 
of stars
in the two brightest magnitudes of MS as function of age. Predictions of our model
for mixed vs unmixed merger are shown on the upper panel, while evolved and unevolved 
are compared on the lower panel.}
\end{figure}

\begin{figure}[t]
\plotone{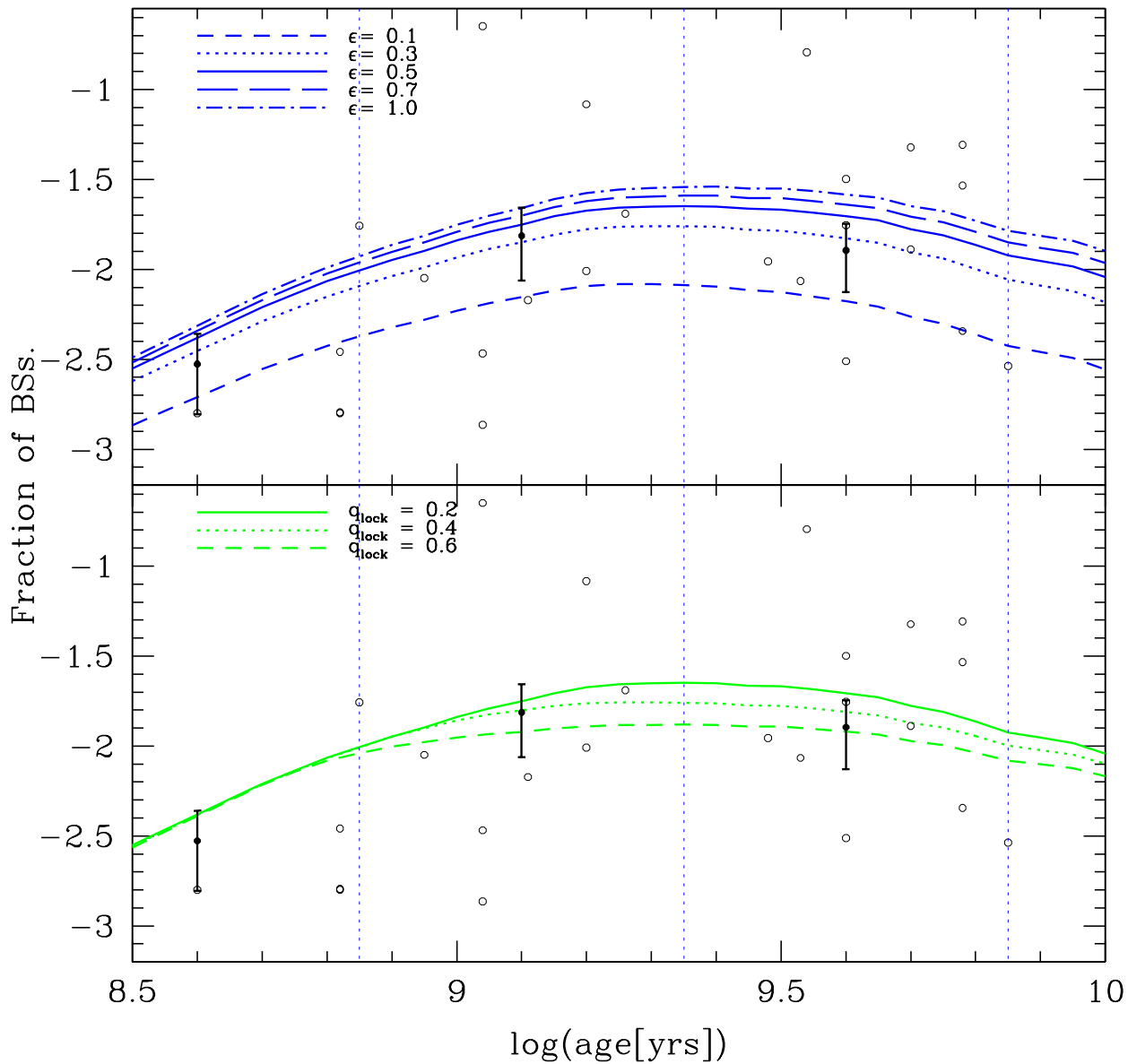}\caption{The predicted number of single blue stragglers normalized to the number 
of stars
in the two brightest magnitudes of MS as a function of age. The upper panel shows the 
predictions of our model for different assumed binary fractions, while models with different 
parameter for tidal locking are compared in the lower panel.}
\end{figure}

\begin{figure}[t]
\plotone{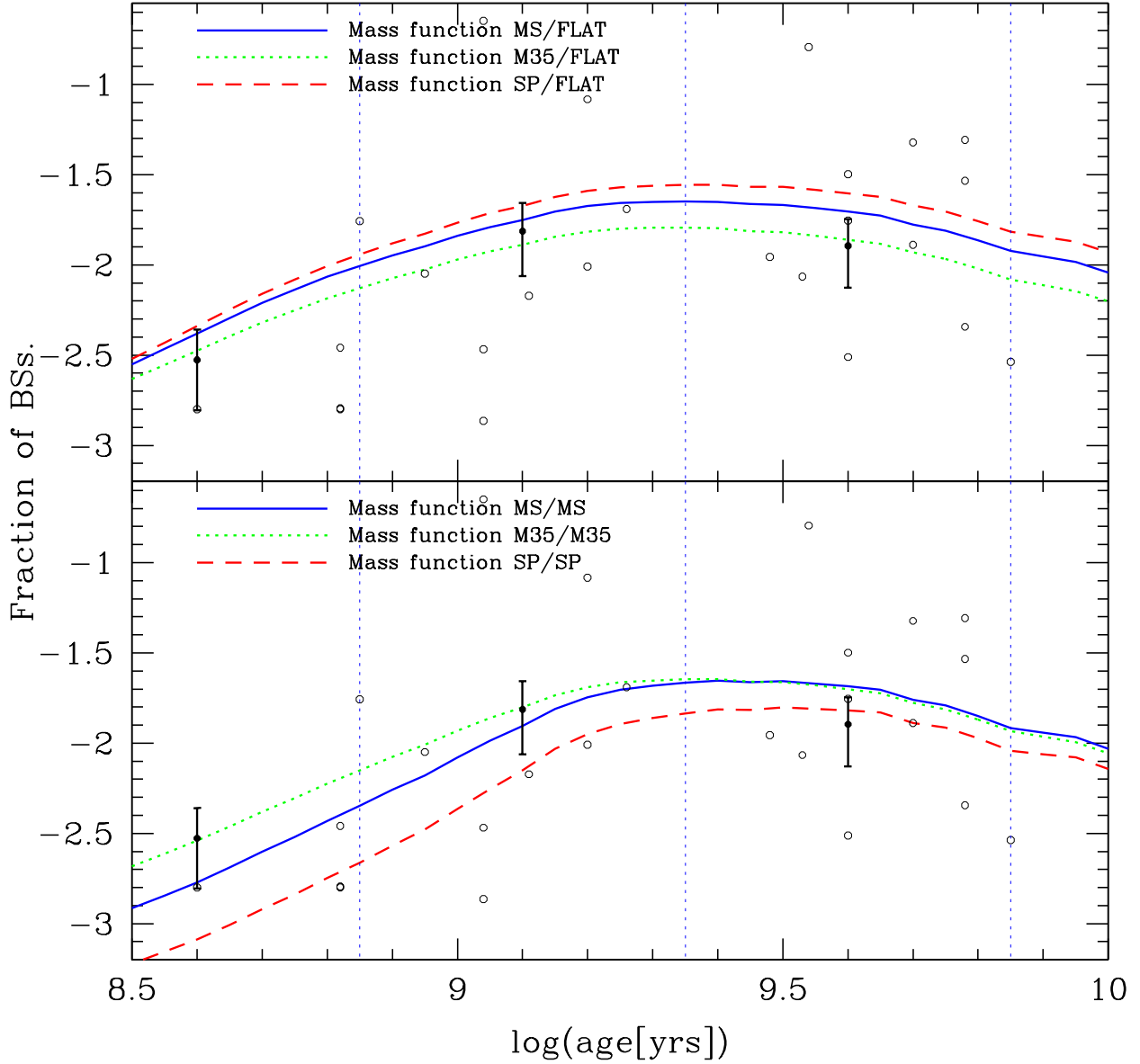}\caption{The predicted number of single blue stragglers normalized to the number 
of stars
in the two brightest magnitudes of MS as function of age. 
Both panels show the predictions of our model for different assumed initial mass functions.
On the upper panel mass fucntion for secondary was assumed to be flat, while on the lower panel
mass functions of primary and secondary are uncorrelated.}
\end{figure}

\begin{figure}[t]
\plotone{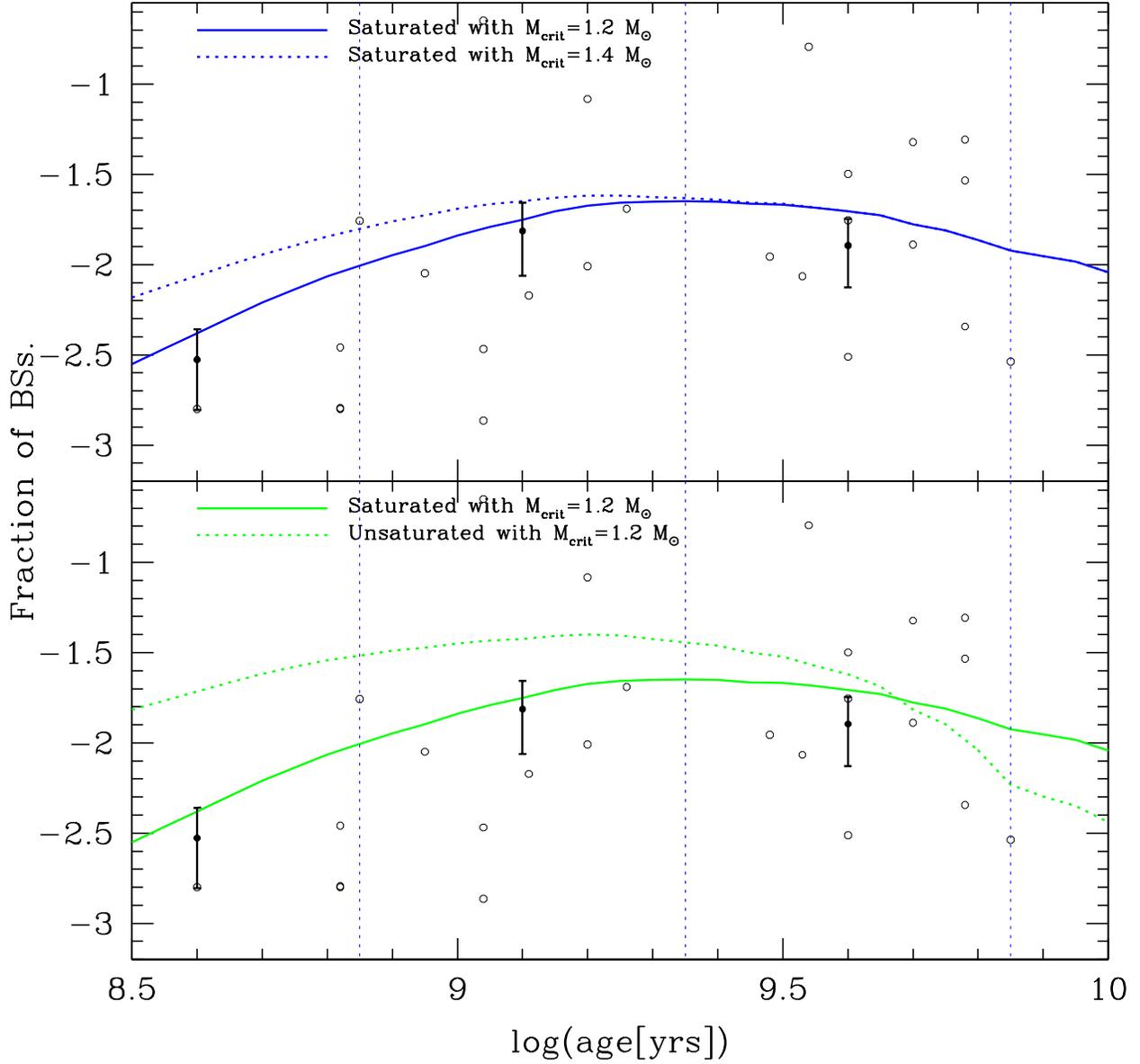}\caption{The predicted number of single blue stragglers normalized to the number 
of stars
in the two brightest magnitudes of MS as function of age. 
Both panels show the predictions of our model for different assumed 
magnetic braking prescription. The lower pannel demonstrates to the
saturated prescription is obviously and unambiquously preferrable then unsaturated one. 
The predictions of the model with the correct magnetic braking prescription but different mass 
limitations for stars experiencing magnetic braking are shown on the upper panel.}
\end{figure}

\begin{figure}[t]
\plotone{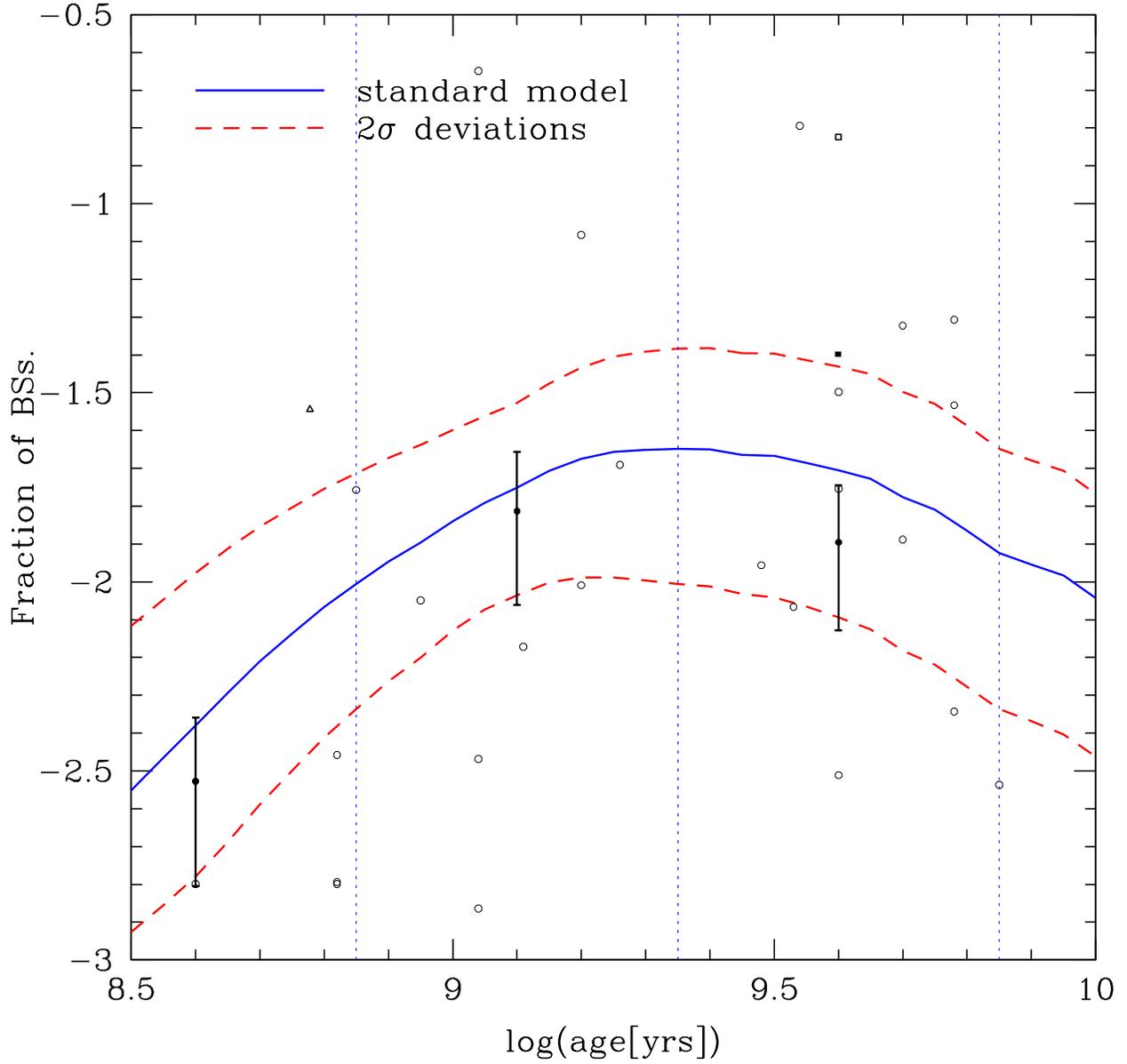}\caption{The number of single blue stragglers normalized to the number of stars
in the two brightest magnitudes of MS as function of age. Our standard model 
is represented by the solid line, while 2-sigma theoretical deviations from 
standard are illustrated by dashed lines.}
\end{figure}

\begin{figure}[t]
\plotone{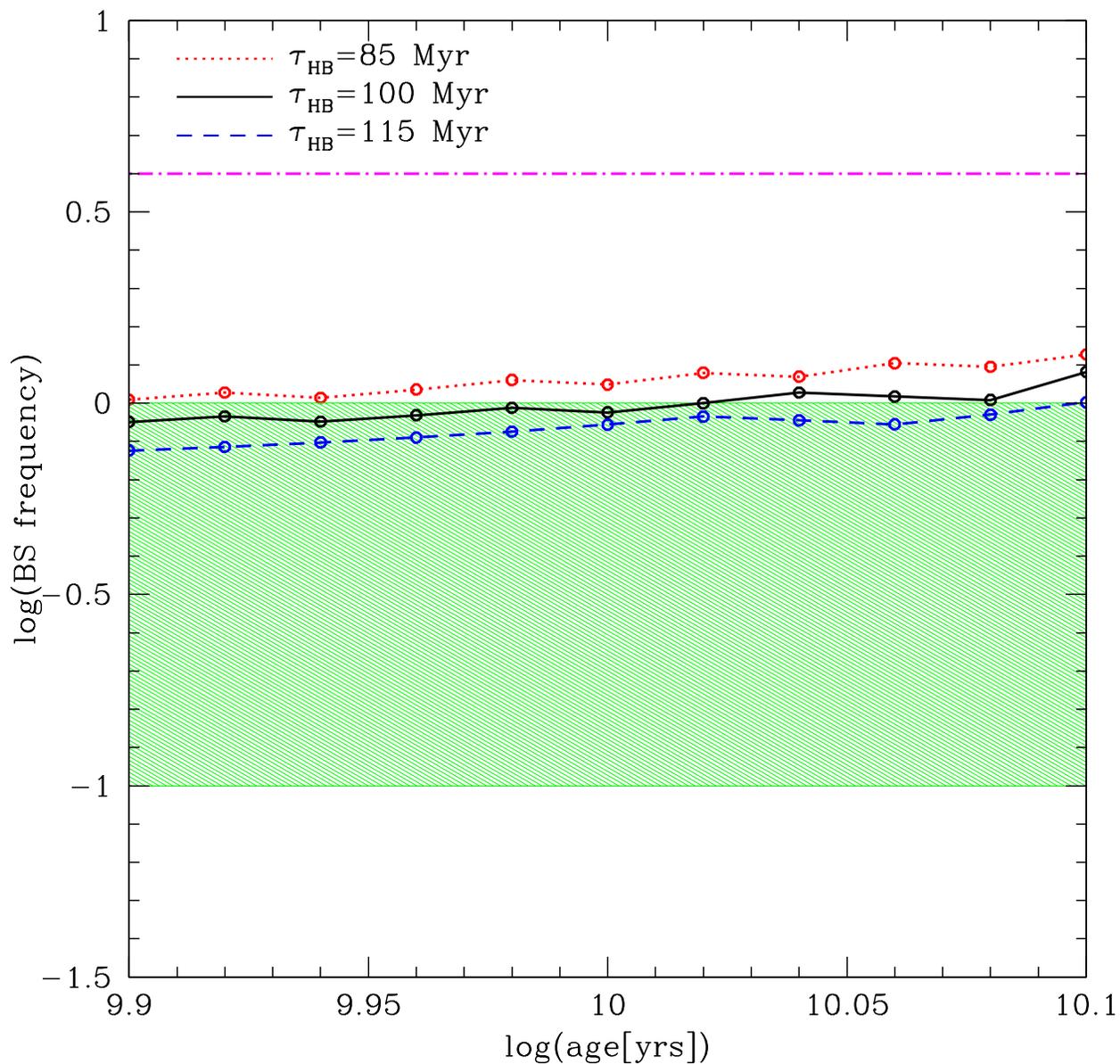}\caption{The predicted number of single blue stragglers normalized to the 
horisontal 
branch stars as function of age. The shaded region denotes the region 
where most globular clusters would be. The dot-dashed line is the approximate frequency
of BMP objects in the galactic halo, while dashed, solid and doted lines show the predictions of our model
withdifferent assumed age on horizontal branch of 85/100/115 Myrs.}
\end{figure}

\begin{figure}[t]
\plotone{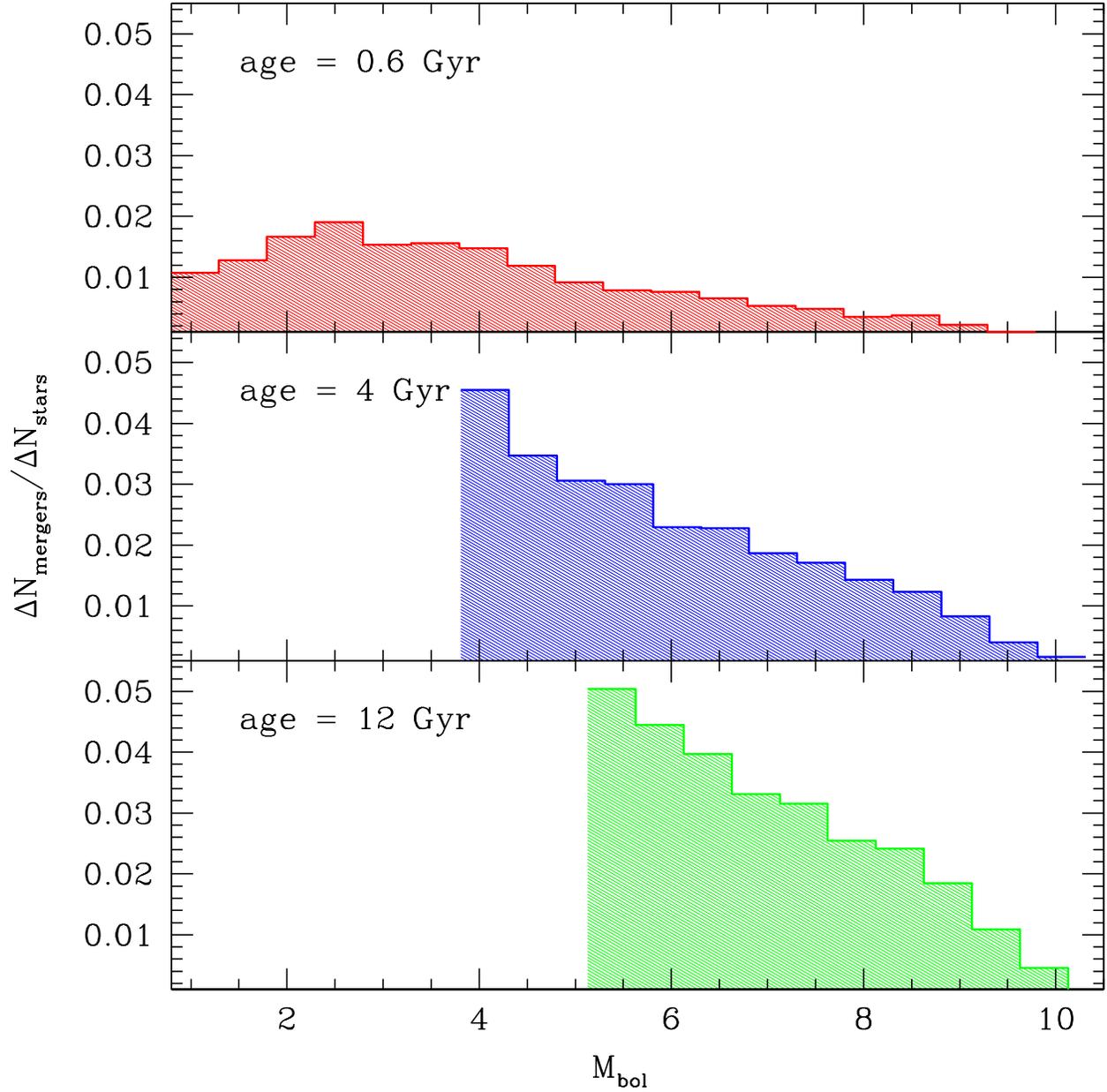}\caption{The predicted fraction of sub turn-off mergers relative to number of MS 
stars at given 
luminosity for three different ages of stellar population. These mergers
probably contribute to lithium-depleted stars.
}
\end{figure}

\begin{figure}[t]
\plotone{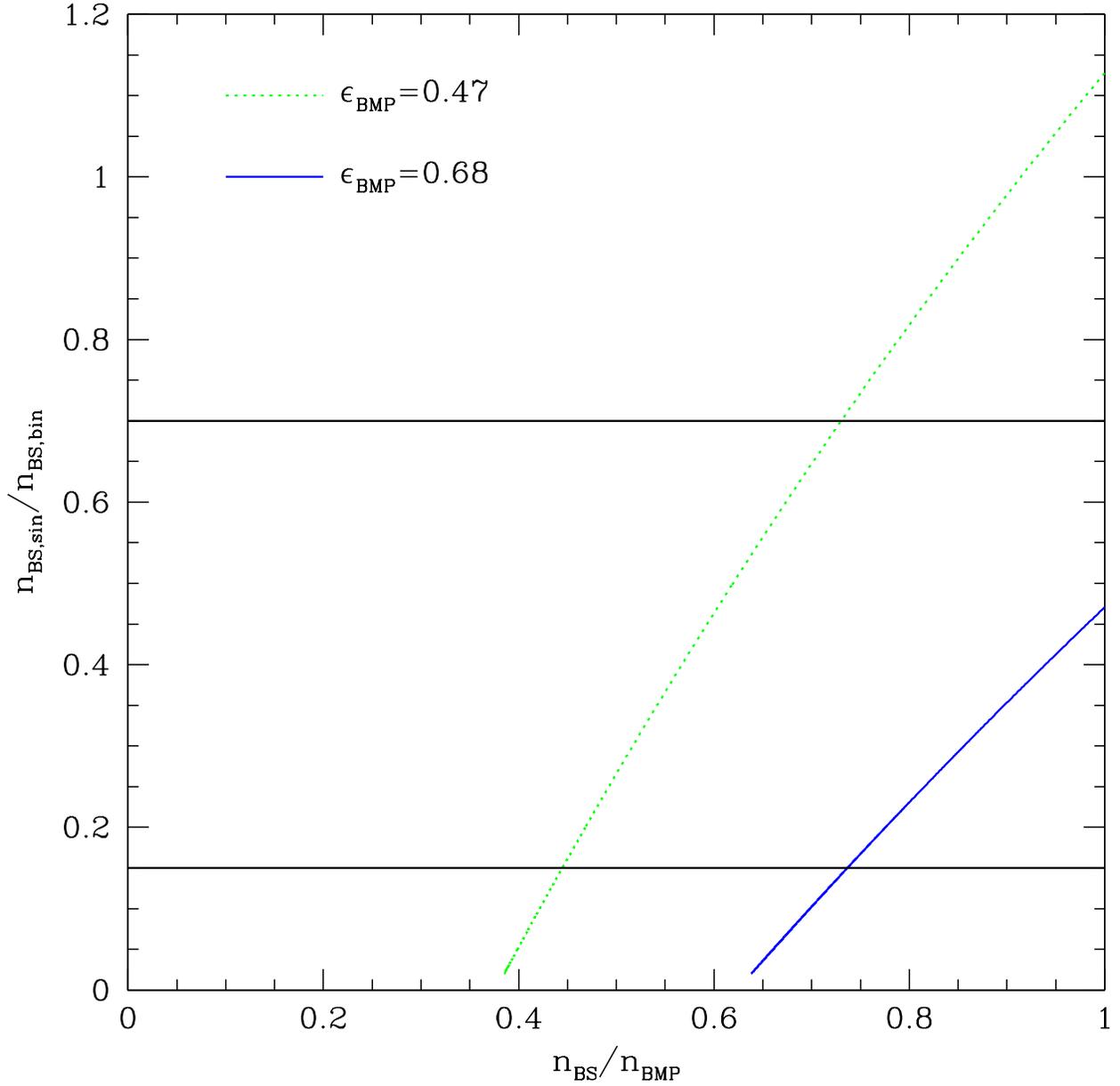}\caption{The relative number of single BSs to BSs found in binaries versus 
fraction of
BSs in the BMP population. Curves show the solution for different assumed binary fractions of 
BMP population. This plot demonstrates our extended order of magnitude estimate 
done by Preston \& Sneden.}
\end{figure}

%% file: tab1.tex
\begin{deluxetable}{rrr} 
\tablecolumns{3} 
\tablewidth{0pc} 
\tablecaption{Parameters of the model} 
\tablehead{ 
\colhead{parameter} & \colhead{values}   & \colhead{section}}
\startdata 
$f(P)$                   & Duquennoy \& Mayor           &  3.1    \\
$\epsilon$               & 0.1 .. 0.6                   &  3.1    \\
$\xi(m_1)$               & $SP$ , $MS$ , $M35$          &  3.1    \\
$\xi(m_2)$               & same as $\xi(m_1)$ or $const$&  3.1    \\
$q_{\rm lock}$               & 0.1 .. 0.5                   &  3.2.1  \\ 
$\dot{J}_{MB}(m,\omega)$ & $saturated$ , $unsaturated$  &  3.2.2  \\
$M_{\rm crit}$               & 1.2 .. 1.4 $M_{\odot}$       &  3.2.2  \\
$product$ $type$         & $mixed$, $unmixed$           &  3.3    \\
$\tau_{\rm cluster}$         & 0.3 .. 12 Gyrs               &  4.1   
\enddata 
\end{deluxetable} 

%% file: tab2.tex
\begin{deluxetable}{rrrrrr}
\tablecolumns{3}
\tablewidth{0pc}
\tablecaption{Parameters of the model}
\tablehead{
\colhead{N} & \colhead{$\epsilon_{BMP}$} & 
\colhead{$\epsilon_{A}$} & \colhead{$\epsilon_{BS}$} & 
\colhead{$n_{BS}/n_{BMP}$} & \colhead{$n_{BS}/n_{HB}$} 
}
\startdata
1 &   0.60        &     0.15       &   0.87          &   0.63           &    4.0           \\
2 &   0.47        &     0.15       &   0.87          &   0.44           &    2.8           \\
3 &   0.60        &     0.15       &   0.65          &   0.90           &    5.7           \\
4 &   0.47        &     0.15       &   0.65          &   0.64           &    4.1           
\enddata
\end{deluxetable}